\pdfoutput=1

\documentclass[12pt,a4paper]{article}

\usepackage{ifthen} 
\newboolean{pdflatex}
\setboolean{pdflatex}{true} 
\usepackage{placeins}
\usepackage{bm}

\newboolean{articletitles}
\setboolean{articletitles}{true} 

\newboolean{uprightparticles}
\setboolean{uprightparticles}{false} 

\def\paperauthors{LHCb collaboration} 
\def\paperasciititle{Measurement of CP observables in B+/- -> D K+/- and B+/- -> D pi+/- decays with D -> Ks0 K pi final states} 
\def\papertitle{Measurement of \CP observables\\ in $\Bpm \to D \Kpm$ and $\Bpm \to D \pipm$ \\with $D \to \KS K^\pm \pi^\mp$ decays} 
\def\paperkeywords{{High Energy Physics}, {LHCb}} 
\def\papercopyright{\the\year\ CERN for the benefit of the LHCb collaboration} 
\def\paperlicence{CC BY 4.0 licence}
\def\paperlicenceurl{https://creativecommons.org/licenses/by/4.0/}

\usepackage[top=1in, bottom=1.25in, left=1in, right=1in]{geometry}

\usepackage{microtype}
\usepackage{lineno}  
\usepackage{xspace} 
\usepackage{caption} 

\usepackage{graphicx}  
\usepackage{color}
\usepackage{colortbl}
\graphicspath{{./figs/}} 

\usepackage{amsmath} 
\usepackage{amssymb}
\usepackage{amsfonts}
\usepackage{upgreek} 

\newcommand*\patchAmsMathEnvironmentForLineno[1]{%
\expandafter\let\csname old#1\expandafter\endcsname\csname #1\endcsname
\expandafter\let\csname oldend#1\expandafter\endcsname\csname
end#1\endcsname
 \renewenvironment{#1}%
   {\linenomath\csname old#1\endcsname}%
   {\csname oldend#1\endcsname\endlinenomath}%
}
\newcommand*\patchBothAmsMathEnvironmentsForLineno[1]{%
  \patchAmsMathEnvironmentForLineno{#1}%
  \patchAmsMathEnvironmentForLineno{#1*}%
}
\AtBeginDocument{%
\patchBothAmsMathEnvironmentsForLineno{equation}%
\patchBothAmsMathEnvironmentsForLineno{align}%
\patchBothAmsMathEnvironmentsForLineno{flalign}%
\patchBothAmsMathEnvironmentsForLineno{alignat}%
\patchBothAmsMathEnvironmentsForLineno{gather}%
\patchBothAmsMathEnvironmentsForLineno{multline}%
\patchBothAmsMathEnvironmentsForLineno{eqnarray}%
}

\usepackage{hyperxmp}

\usepackage[pdftex,
            pdfauthor={\paperauthors},
            pdftitle={\paperasciititle},
            pdfkeywords={\paperkeywords},
            pdfcopyright={Copyright (C) \papercopyright},
            pdflicenseurl={\paperlicenceurl}]{hyperref}

\usepackage[colorinlistoftodos,textsize=scriptsize]{todonotes}

\usepackage[bottom,flushmargin,hang,multiple]{footmisc}

\usepackage[all]{hypcap} 

\usepackage{xspace} 
\usepackage{upgreek}

\def\lhcb   {\mbox{LHCb}\xspace}

\def\MagUp {\mbox{\em Mag\kern -0.05em Up}\xspace}

\ifthenelse{\boolean{uprightparticles}}%
{
 
 \def\Pgamma      {\ensuremath{\upgamma}\xspace}

 \def\Ppi         {\ensuremath{\uppi}\xspace}

 \def\PDelta      {\ensuremath{\Delta}\xspace}                 
 \def\PXi         {\ensuremath{\Xi}\xspace}                 
 \def\PLambda     {\ensuremath{\Lambda}\xspace}                 
 \def\PSigma      {\ensuremath{\Sigma}\xspace}                 
 \def\POmega      {\ensuremath{\Omega}\xspace}                 
 \def\PUpsilon    {\ensuremath{\Upsilon}\xspace}

 \def\PB      {\ensuremath{\mathrm{B}}\xspace}                 
                  
 \def\PD      {\ensuremath{\mathrm{D}}\xspace}

 \def\PK      {\ensuremath{\mathrm{K}}\xspace}

 \def\Pb      {\ensuremath{\mathrm{b}}\xspace}                 
 \def\Pc      {\ensuremath{\mathrm{c}}\xspace}

 \def\Pi      {\ensuremath{\mathrm{i}}\xspace}

 \def\Ps      {\ensuremath{\mathrm{s}}\xspace}

 \def\thebaroffset{0.0em}
}
{
 
 \def\Pgamma      {\ensuremath{\gamma}\xspace}

 \def\Ppi         {\ensuremath{\pi}\xspace}

 \mathchardef\PDelta="7101
 \mathchardef\PXi="7104
 \mathchardef\PLambda="7103
 \mathchardef\PSigma="7106
 \mathchardef\POmega="710A
 \mathchardef\PUpsilon="7107
                  
 \def\PB      {\ensuremath{B}\xspace}                 
                  
 \def\PD      {\ensuremath{D}\xspace}

 \def\PK      {\ensuremath{K}\xspace}

 \def\Pb      {\ensuremath{b}\xspace}                 
 \def\Pc      {\ensuremath{c}\xspace}

 \def\Pi      {\ensuremath{i}\xspace}

 \def\Ps      {\ensuremath{s}\xspace}

 \def\thebaroffset{0.18em}
}
\newcommand{\offsetoverline}[2][\thebaroffset]{\kern #1\overline{\kern -#1 #2}}%

\makeatletter
\ifcase \@ptsize \relax
  \newcommand{\miniscule}{\@setfontsize\miniscule{4}{5}}
\or
  \newcommand{\miniscule}{\@setfontsize\miniscule{5}{6}}
\or
  \newcommand{\miniscule}{\@setfontsize\miniscule{5}{6}}
\fi
\makeatother

\DeclareRobustCommand{\optbar}[1]{\shortstack{{\miniscule (\rule[.5ex]{1.25em}{.18mm})}
  \\ [-.7ex] $#1$}}

\def\g      {{\ensuremath{\Pgamma}}\xspace}

\def\squark    {{\ensuremath{\Ps}}\xspace}

\def\cquark    {{\ensuremath{\Pc}}\xspace}

\def\bquark    {{\ensuremath{\Pb}}\xspace}

\def\pion   {{\ensuremath{\Ppi}}\xspace}
\def\piz    {{\ensuremath{\pion^0}}\xspace}
\def\pip    {{\ensuremath{\pion^+}}\xspace}
\def\pim    {{\ensuremath{\pion^-}}\xspace}
\def\pipm   {{\ensuremath{\pion^\pm}}\xspace}
\def\pimp   {{\ensuremath{\pion^\mp}}\xspace}

\def\kaon    {{\ensuremath{\PK}}\xspace}

\def\KorKbar {\kern \thebaroffset\optbar{\kern -\thebaroffset \PK}{}\xspace}

\def\Kp      {{\ensuremath{\kaon^+}}\xspace}
\def\Km      {{\ensuremath{\kaon^-}}\xspace}
\def\Kpm     {{\ensuremath{\kaon^\pm}}\xspace}
\def\Kmp     {{\ensuremath{\kaon^\mp}}\xspace}
\def\KS      {{\ensuremath{\kaon^0_{\mathrm{S}}}}\xspace}

\def\Kstar   {{\ensuremath{\kaon^*}}\xspace}

\def\Kstarp  {{\ensuremath{\kaon^{*+}}}\xspace}

\def\Dbar    {{\ensuremath{\offsetoverline{\PD}}}\xspace}
\def\D       {{\ensuremath{\PD}}\xspace}

\def\DorDbar {\kern \thebaroffset\optbar{\kern -\thebaroffset \PD}\xspace}
\def\Dz      {{\ensuremath{\D^0}}\xspace}
\def\Dzb     {{\ensuremath{\Dbar{}^0}}\xspace}
\def\Dp      {{\ensuremath{\D^+}}\xspace}

\def\Dstarz  {{\ensuremath{\D^{*0}}}\xspace}

\def\B       {{\ensuremath{\PB}}\xspace}

\def\BorBbar {\kern \thebaroffset\optbar{\kern -\thebaroffset \PB}\xspace}
\def\Bz      {{\ensuremath{\B^0}}\xspace}

\def\Bd      {{\ensuremath{\B^0}}\xspace}

\def\BdorBdbar {\kern \thebaroffset\optbar{\kern -\thebaroffset \Bd}\xspace}
\def\Bu      {{\ensuremath{\B^+}}\xspace}
\def\Bub     {{\ensuremath{\B^-}}\xspace}
\def\Bp      {{\ensuremath{\Bu}}\xspace}
\def\Bm      {{\ensuremath{\Bub}}\xspace}
\def\Bpm     {{\ensuremath{\B^\pm}}\xspace}

\def\Bs      {{\ensuremath{\B^0_\squark}}\xspace}

\def\BsorBsbar {\kern \thebaroffset\optbar{\kern -\thebaroffset \Bs}\xspace}

\def\Y#1S{\ensuremath{\PUpsilon{(#1S)}}\xspace}

\def\LorLbar     {\kern \thebaroffset\optbar{\kern -\thebaroffset \PLambda}\xspace}

\def\to                 {\ensuremath{\rightarrow}\xspace}

\def\CP                {{\ensuremath{C\!P}}\xspace}

\def\AT#1     {\ensuremath{A_{\mathrm{T}}^{#1}}\xspace}           

\def\C#1      {\ensuremath{\mathcal{C}_{#1}}\xspace}                       
\def\Cp#1     {\ensuremath{\mathcal{C}_{#1}^{'}}\xspace}                    
\def\Ceff#1   {\ensuremath{\mathcal{C}_{#1}^{\mathrm{(eff)}}}\xspace}        
\def\Cpeff#1  {\ensuremath{\mathcal{C}_{#1}^{'\mathrm{(eff)}}}\xspace}       
\def\Ope#1    {\ensuremath{\mathcal{O}_{#1}}\xspace}                       
\def\Opep#1   {\ensuremath{\mathcal{O}_{#1}^{'}}\xspace}                    

\newcommand{\nospaceunit}[1]{\ensuremath{\text{#1}}}       
\newcommand{\aunit}[1]{\ensuremath{\text{\,#1}}}       

\newcommand{\tev}{\aunit{Te\kern -0.1em V}\xspace}
\newcommand{\gev}{\aunit{Ge\kern -0.1em V}\xspace}
\newcommand{\mev}{\aunit{Me\kern -0.1em V}\xspace}
\newcommand{\kev}{\aunit{ke\kern -0.1em V}\xspace}
\newcommand{\ev}{\aunit{e\kern -0.1em V}\xspace}
\newcommand{\mevc}{\ensuremath{\aunit{Me\kern -0.1em V\!/}c}\xspace}
\newcommand{\gevc}{\ensuremath{\aunit{Ge\kern -0.1em V\!/}c}\xspace}
\newcommand{\mevcc}{\ensuremath{\aunit{Me\kern -0.1em V\!/}c^2}\xspace}
\newcommand{\gevcc}{\ensuremath{\aunit{Ge\kern -0.1em V\!/}c^2}\xspace}

\def\mum  {\ensuremath{\,\upmu\nospaceunit{m}}\xspace}

\def\fb   {\ensuremath{\aunit{fb}}\xspace}
\def\invfb   {\ensuremath{\fb^{-1}}\xspace}

\newcommand{\chisq}{\ensuremath{\chi^2}\xspace}

\newcommand{\chisqip}{\ensuremath{\chi^2_{\text{IP}}}\xspace}

\def\gsim{{~\raise.15em\hbox{$>$}\kern-.85em
          \lower.35em\hbox{$\sim$}~}\xspace}
\def\lsim{{~\raise.15em\hbox{$<$}\kern-.85em
          \lower.35em\hbox{$\sim$}~}\xspace}

\def\sqs   {\ensuremath{\protect\sqrt{s}}\xspace}

\def\pt         {\ensuremath{p_{\mathrm{T}}}\xspace}

\def\ptot       {\ensuremath{p}\xspace}

\def\evtgen     {\mbox{\textsc{EvtGen}}\xspace}

\def\geant      {\mbox{\textsc{Geant4}}\xspace}

\def\photos     {\mbox{\textsc{Photos}}\xspace}

\def\pythia     {\mbox{\textsc{Pythia}}\xspace}

\def\tell1  {TELL1\xspace}
\def\ukl1   {UKL1\xspace}

\usepackage{cite} 
\usepackage{mciteplus}

\usepackage{longtable} 

\begin{document}

\renewcommand{\thefootnote}{\fnsymbol{footnote}}
\setcounter{footnote}{1}


\begin{titlepage}
\pagenumbering{roman}

\vspace*{-1.5cm}
\centerline{\large EUROPEAN ORGANIZATION FOR NUCLEAR RESEARCH (CERN)}
\vspace*{1.5cm}
\noindent
\begin{tabular*}{\linewidth}{lc@{\extracolsep{\fill}}r@{\extracolsep{0pt}}}
\ifthenelse{\boolean{pdflatex}}
{\vspace*{-1.5cm}\mbox{\!\!\!\includegraphics[width=.14\textwidth]{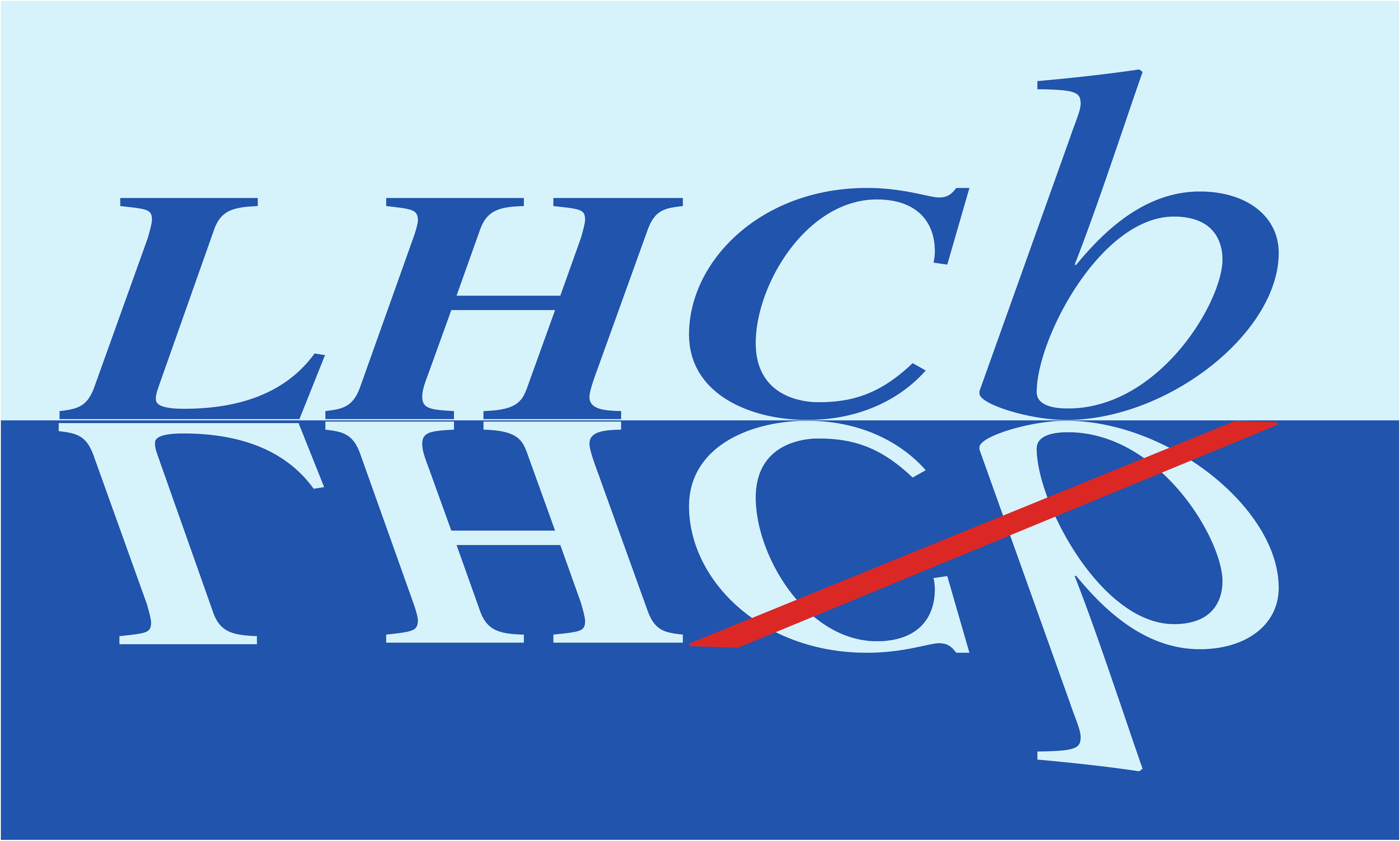}} & &}%
{\vspace*{-1.2cm}\mbox{\!\!\!\includegraphics[width=.12\textwidth]{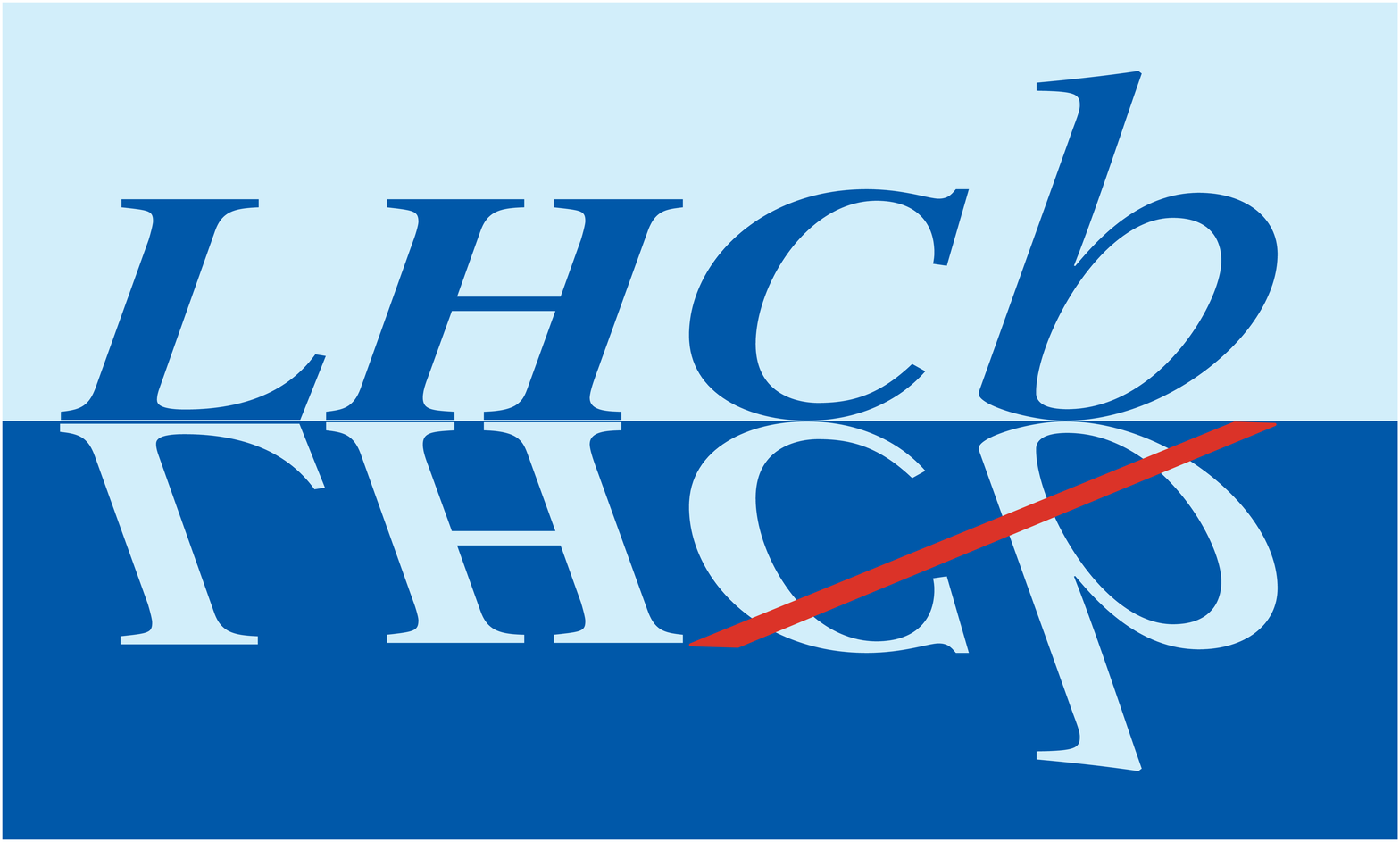}} & &}%
\\
 & & CERN-EP-2020-015 \\  
 & & LHCb-PAPER-2019-044 \\  
 & & \today \\ 
 & & \\
\end{tabular*}

\vspace*{4.0cm}

{\normalfont\bfseries\boldmath\huge
\begin{center}
  \papertitle 
\end{center}
}

\vspace*{2.0cm}

\begin{center}
\paperauthors\footnote{Authors are listed at the end of this paper.}
\end{center}

\vspace{\fill}

\begin{abstract}
  \noindent
  Measurements of \CP observables in $\Bpm \to D \Kpm$ and $\Bpm \to D \pipm$ decays are presented, where $D$ represents a superposition of $\Dz$ and $\Dzb$ states. The $D$ meson is reconstructed in the three-body final states $\KS\Kpm\pimp$ and $\KS\Kmp\pipm$. The analysis uses samples of $B$ mesons produced in proton-proton collisions, corresponding to an integrated luminosity of 1.0, 2.0, and 6.0\invfb\ collected with the \lhcb\ detector at centre-of-mass energies of $\sqs = $ 7, 8, and 13\tev, respectively. These measurements are the most precise to date, and provide important input for the determination of the CKM angle $\gamma$. 
\end{abstract}

\vspace*{2.0cm}

\begin{center}
  Published in J.~High Energ.~Phys.~2020, 58 (2020)
\end{center}

\vspace{\fill}

{\footnotesize 
\centerline{\copyright~\papercopyright. \href{\paperlicenceurl}{\paperlicence}.}}
\vspace*{2mm}

\end{titlepage}


\newpage
\setcounter{page}{2}
\mbox{~}
%
%
%
%

\cleardoublepage


\renewcommand{\thefootnote}{\arabic{footnote}}
\setcounter{footnote}{0}



\pagestyle{plain} 
\setcounter{page}{1}
\pagenumbering{arabic}


%

\section{Introduction}
\label{sec:Introduction}

In the Standard Model (SM), \CP violation in the hadronic sector is described by the irreducible complex phase of the Cabibbo--Kobayashi--Maskawa (CKM) quark mixing matrix\cite{Cabibbo:1963yz,Kobayashi:1973fv}. This matrix is unitary, which leads to the condition $V_{ud}V^*_{ub} + V_{cd}V^*_{cb} + V_{td}V^*_{tb} = 0$, where $V_{ij}$ is the CKM matrix element relating quark $i$ to quark $j$. This condition can be represented by a triangle in the complex plane with internal angles $\alpha$, $\beta$, and $\gamma$. The angle $\gamma$ is defined as $\gamma \equiv \arg{(-{V_{ud}V^*_{ub}}/{V_{cd}V^*_{cb}})}$, which is equal to $\arg{(-{V_{us}V^*_{ub}}/{V_{cs}V^*_{cb}})}$ up to $\mathcal{O}(\lambda^4) \sim 10^{-3}$ in the Wolfenstein parameterisation\cite{wolfenstein}, where $\lambda$ is the sine of the Cabibbo angle~\cite{Cabibbo:1963yz}. Improving knowledge of $\gamma$ can be achieved in a theoretically clean manner by studying the interference of  $b \to u$ and $b \to c$ transition amplitudes in tree-level $b$-hadron decays. Such a measurement provides a benchmark against which other flavour observables that are more susceptible to the influence of physics beyond the SM can be compared~\cite{King:2019rvk}.

A combination of measurements from LHCb currently yields $\gamma = (74.0\,^{+5.0}_{-5.8})^{\circ}$~\cite{LHCb-CONF-2018-002,LHCb-PAPER-2016-032}, which is the most precise determination of $\gamma$ from a single experiment. The precision is dominated by measurements exploiting the $\Bp \to D \Kp$ decay,\footnote{The inclusion of charge-conjugate processes is implied throughout, unless otherwise indicated.} where $D$ indicates a superposition of $\Dz$ and $\Dzb$ mesons reconstructed in a final state common to both. To continue improving the precision on $\gamma$, independent measurements can be performed using all suitable $D$ meson final states. Several different final states have thus far been analysed at LHCb, including a previous measurement of the singly Cabibbo-suppressed $D \to \KS \Km \pip$ and $D \to \KS \Kp \pim$ modes~\cite{LHCb-PAPER-2013-068}. These decays are reconstructed in two categories by comparing the charge of the pion produced in the $D$ decay with the charge of the $B$ meson; $\Bp \to [\KS \Kp \pim]_D h^+$ decays are thus labelled ``Same Sign'' (SS), and \mbox{$\Bp \to [\KS \Km \pip]_D h^+$} decays are labelled ``Opposite Sign'' (OS), where $h \in \{\pi,K\}$.
This paper reports an update to Ref.~\cite{LHCb-PAPER-2013-068}, measuring \CP observables in $\Bp \to D \Kp$ and $\Bp \to D \pip$ decays using the $D\to \KS \Kp \pim$ and $D\to \KS \Km \pip$ final states. 
Data corresponding to 6.0\invfb\ of integrated luminosity collected between 2015 and 2018 (Run 2) of data taking is used. The Run 1 dataset collected during 2011 and 2012 and corresponding to an integrated luminosity of 3.0\invfb\ is also reprocessed, to benefit from an improved selection as well as a reappraisal of the backgrounds.

In order to interpret interference effects involving multi-body $D$-decays, it is necessary to account for the amplitude structure of the Dalitz plot. Instead of employing an amplitude model to describe the contributing partial waves, the CLEO collaboration have made measurements of the effective amplitude and phase variation using a sample of quantum-correlated $D$ decays collected by the CLEO-c experiment~\cite{CLEOKsKPi}. Due to limited sample size, those measurements were performed averaging over large regions of the Dalitz plot, notably defining one of two regions to contain the $D\to \Kstar(892)^+\Km$ mode.  In the present work, results are reported for both the \Kstarp and non-\Kstarp regions of the Dalitz plot, respecting the boundary defined by CLEO-c. The use of external CLEO-c results, which were performed across the full Dalitz plot and within the $K^{*+}$ region, avoids the need to introduce a systematic uncertainty resulting from an amplitude model description.

The paper is organised as follows:  Sec.~\ref{sec:formalism} presents the observables to be measured and their relationships to the physics parameters of interest;  Sec.~\ref{sec:Detector} discusses the aspects of the detector, trigger, and simulation that are relevant for the measurement; Secs.~\ref{sec:selection},~\ref{sec:mass_fit}, and~\ref{sec:systematics} describe the candidate selection, the fit to the invariant mass spectra, and the assignment of systematic uncertainties;  the observable results are presented in Sec.~\ref{sec:results}.

\section{Formalism}
\label{sec:formalism}
The SS $\Bp \to [\KS\Kp \pim]_D \Kp$ decay can proceed via the \Dz or \Dzb states. As such, the total decay amplitude is given by the sum of two interfering amplitudes,
\begin{eqnarray}
A_{\KS\Kp\pim}(\bm{x}) = A_{\Dz}(\bm{x})&  +& r_B e^{i(\delta_B + \gamma)} A_{\Dzb}(\bm{x}),
\end{eqnarray}
where $\bm{x}$ represents the Dalitz plot coordinates $(m^2_{\KS K}, m^2_{\KS\pi})$, $A_{\{\Dz,\Dzb\}}(\bm{x})$ are the $\Dz$ and $\Dzb$ decay amplitudes at a specific point in the $\KS\Kp\pim$ Dalitz plot~\cite{Grossman:2002aq}. The OS $\Bp \to [\KS\Km \pip]_D \Kp$ decay also proceeds via both \Dz and \Dzb, with a total decay amplitude given by
\begin{eqnarray}
A_{\KS\Km\pip}(\bm{x}) = A_{\Dzb}(\bm{x})&  +& r_B e^{i(\delta_B + \gamma)} A_{\Dz}(\bm{x})\,.
\end{eqnarray}
The amplitude ratio $r_B = \frac{| A(\Bp \to \Dz \Kp) |}{| A(\Bp \to \Dzb \Kp) |}\sim 0.1$~\cite{LHCb-CONF-2018-002,LHCb-PAPER-2016-032}, and $\delta_B = \arg\left( \frac{A(\Bp \to \Dz \Kp)}{A(\Bp \to \Dzb \Kp)}  \right)$ is the strong-phase difference between the $B$ decay amplitudes. To calculate the decay rate in a finite region of the Dalitz plot, the integral of the interference term over that region must be known. 
In Ref.~\cite{CLEOKsKPi}, measurements of quantum-correlated $D$ decays have been used to determine the amplitude ratio, $r_D = \frac{| A(\Dz \to K_s^0 \Kp \pim) |}{| A(\Dzb \to K_s^0 \Kp \pim)|}$, and the integral of the interference term directly in the form of a coherence factor, $\kappa_D$, and an average strong phase difference, $\delta_D$~\cite{Atwood:2003mj}. The coherence factor is defined as
\begin{align}
\small
\kappa_{D}\, e^{-i\delta_{D}} = \frac{\int A^*_{K_S^0K^-\pi^+}(\bm{x})\, A_{K_S^0K^+\pi^-} (\bm{x})\, d\bm{x}}{\sqrt{\int |A_{K_S^0K^-\pi^+}(\bm{x})|^2 d\bm{x}} \hspace{0.1cm} \sqrt{\int |A_{K_S^0K^+\pi^-}(\bm{x})|^2 d\bm{x}}}\,. \label{eq:defCohFac}
\end{align}
A similar notation also holds for SS and OS $B^+ \to D\pip$ decays with the replacements $r_B \to r_B^\pi$ and $\delta_B \to \delta_B^\pi$, where $r_{B}^{\pi} \sim 0.015$.

In each Dalitz region, four decay rates are considered in this analysis~\cite{Atwood:1996ci}:
\begin{equation}
\begin{aligned}
N_{\rm{SS}}^{DK^\pm} &\propto& 1 + r_{B}^2r_D^2 + 2 r_{B}r_D\kappa_{D}\cos(\delta_{B} \pm \gamma - \delta_{D})\,, \\
N_{\rm{OS}}^{DK^\pm} &\propto& r_{B}^2 + r_D^2 + 2 r_{B}r_D\kappa_{D}\cos(\delta_{B} \pm \gamma + \delta_{D})\,,\\
N_{\rm{SS}}^{D\pi^\pm} &\propto& 1 + (r_{B}^{\pi})^{2}r_D^2 + 2 r_{B}^{\pi}r_D\kappa_{D}\cos(\delta_{B}^{\pi} \pm \gamma - \delta_{D})\,, \\
N_{\rm{OS}}^{D\pi^\pm} &\propto& (r_{B}^{\pi})^{2} + r_D^2 + 2 r_{B}^{\pi}r_D\kappa_{D}\cos(\delta_{B}^{\pi} \pm \gamma + \delta_{D})\,.
\label{eqn:defineYieldsGammetal}
\end{aligned}
\end{equation}
Observables constructed from Eq. \ref{eqn:defineYieldsGammetal} have sensitivity to $\gamma$ that depends upon the value of the coherence factor, with a higher coherence corresponding to greater sensitivity. The CLEO-c results~\cite{CLEOKsKPi} show high coherence within the \Kstarp region, defined as $\pm 100\mevcc$ around the \Kstarp mass; $\kappa_{D} = 0.94 \pm 0.12$ and $\delta_{D} = (-16.6 \pm 18.4)^\circ$ are reported. 
With $r_D\approx0.6$~\cite{CLEOKsKPi}, the maximal \CP asymmetry that can be expected is $35\%$ in $\Bp\to D\Kp$ decays, but only $2\%$ in $\Bp\to D\pip$ decays due to the dissimilarity of $r_D$ and $r_{B}^{\pi}$. 
Dedicated measurements in the non-\Kstarp region have not yet been made.
Eight yields are measured in this analysis, from which seven ratios are constructed as \CP observables; each observable can be related to $\gamma$ through the decay rates in Eq.~\ref{eqn:defineYieldsGammetal}.
The charge asymmetry is measured in four decay modes,
\begin{equation*}
A_{m}^{Dh} = \frac{N_{m}^{Dh^-} - N_{m}^{Dh^+}}{N_{m}^{Dh^-} + N_{m}^{Dh^+}}\,,
\end{equation*}
where $m \in \{\text{SS,OS}\}$ and $h \in \{\pi,K\}$. The ratios of $\Bp \to D \Kp$ and $\Bp \to D \pip$ yields, $R_{m}^{DK/D\pi}$, are determined, and the ratio of SS to OS $\Bp \to D \pip$ yields, $R_{SS/OS}$, is also measured. The measurements are reported for the \Kstarp region of the $D$ Dalitz plot as defined above, and outside it; they are not interpreted in terms of $\gamma$ in this work, as constraints on the $B$ decay hadronic parameters which come from measurements using other $D$ decay modes are necessary at the current level of statistical precision.
\section{Detector and simulation}
\label{sec:Detector}
The \lhcb detector~\cite{LHCb-DP-2008-001,LHCb-DP-2014-002} is a single-arm forward
spectrometer covering the \mbox{pseudorapidity} range $2<\eta <5$,
designed for the study of particles containing \bquark or \cquark
quarks. The detector includes a high-precision tracking system
consisting of a silicon-strip vertex detector surrounding the $pp$
interaction region, a silicon-strip detector located
upstream of a dipole magnet with a bending power of about
$4{\mathrm{\,Tm}}$, and three stations of silicon-strip detectors and straw
drift tubes placed downstream of the magnet.
The tracking system provides a measurement of the momentum, \ptot, of charged particles with relative uncertainty that varies from 0.5\% at low momentum to 1.0\% at 200\gevc.
The minimum distance of a track to a primary vertex (PV), the impact parameter (IP), 
is measured with a resolution of $(15+29/\pt)\mum$,
where \pt is the component of the momentum transverse to the beam, in\,\gevc.
Different types of charged hadrons are distinguished using information
from two ring-imaging Cherenkov (RICH) detectors. 
Photons, electrons and hadrons are identified by a calorimeter system consisting of
scintillating-pad and preshower detectors, an electromagnetic
and a hadronic calorimeter. Muons are identified by a
system composed of alternating layers of iron and multiwire
proportional chambers.

The online event selection is performed by a trigger, 
which consists of a hardware stage, based on information from the calorimeter and muon
systems, followed by a software stage, which applies a full event
reconstruction. The events considered in the analysis are triggered at the hardware level when either one of the final-state tracks of the signal decay deposits enough energy in the calorimeter system, or when one of the other particles in the event, not reconstructed as part of the signal candidate, fulfils any trigger requirement. 
At the software stage, it is required that at least one particle should have high \pt\ and high \chisqip, where \chisqip\ is defined as the difference in the PV fit \chisq\ with and without the inclusion of that particle. A multivariate algorithm~\cite{BBDT} is used to identify secondary vertices consistent with being a two-, three-, or four-track $b$-hadron decay. The PVs are fitted with and without the $B$ candidate tracks, and the PV that gives the smallest \chisqip\ is associated with the $B$ candidate.

Simulated events are used to describe the signal mass shapes and compute efficiencies. In the simulation, $pp$ collisions are generated using
\pythia~\cite{Sjostrand:2007gs} with a specific \lhcb\ configuration~\cite{LHCb-PROC-2010-056}. Decays of hadronic particles
are described by \evtgen~\cite{Lange:2001uf}, in which final-state
radiation is generated using \photos~\cite{Golonka:2005pn}. The
interaction of the generated particles with the detector, and its response,
are implemented using the \geant
toolkit~\cite{Allison:2006ve, *Agostinelli:2002hh} as described in
Ref.~\cite{LHCb-PROC-2011-006}.

\section{Offline selection}
\label{sec:selection}

Decays of \KS mesons to the $\pip\pim$ final state are reconstructed in two categories,
the first containing \KS mesons that decay early enough for the pions to be reconstructed in the vertex detector, and the
second containing \KS mesons that decay later such that track segments of the pions cannot be formed in the vertex detector. These categories are
referred to as \emph{long} and \emph{downstream}, respectively. The candidates in the long category have better mass, momentum, and vertex resolution 
than those in the downstream category, but the downstream category contains more candidates and thus both are used. Herein, \Bp candidates are denoted long or downstream depending on which category of \KS candidate is used.

The $D$ ($\KS$) candidates are required to be within $\pm25\mevcc$ ($\pm15\mevcc$) of the known mass~\cite{PDG2018}, and \Bp meson candidates with invariant masses in the interval \mbox{5080--5700\mevcc} are retained. The kaons and pions originating from both the \Bp and \D decays are required to have \pt in the range 0.5--10\gevc and $p$ in the range 5--100\gevc. 
These requirements ensure that the tracks are within the kinematic coverage of the RICH detectors, which are used to provide particle identification (PID) information. 

A boosted decision tree (BDT) classifier~\cite{Breiman} implementing the gradient boost algorithm is employed to achieve further combinatorial background suppression. The BDT is trained using simulated \mbox{$\Bp \to D h^+$} decays as a proxy for signal and a background sample of candidates in data with invariant masses in the range 5900--7200\mevcc which are not used in the invariant-mass fit (see Sec.~\ref{sec:mass_fit}). The input to the BDT is a set of features that characterise the signal decay. These features can be divided into two categories:
(1) properties of any particle, and (2) properties of composite particles only (the \D and \Bp candidates). Specifically:
\begin{enumerate}
\item{$p$, \pt, and \chisqip;}
\item{decay time, flight distance between production and decay vertex, decay vertex quality, radial distance between the decay vertex and the PV, and the angle between the particle's momentum vector and the line connecting the production and decay vertices.}
\end{enumerate} 
In addition, a feature that estimates the imbalance of \pt around the \Bp candidate momentum vector is also used in the BDT. It is defined as
\begin{equation}
I_{\pt} = \frac{\pt(\Bp) - \Sigma \pt}{\pt(\Bp) + \Sigma \pt}\,,
\end{equation}
where the sum is taken over tracks inconsistent with originating from the PV that lie within a cone around the \Bp candidate, excluding tracks used to make the signal candidate.
The cone is defined by a circle with a radius of 1.5 units in the plane of pseudorapidity and azimuthal angle expressed in radians. Including the $I_{\pt}$ feature in the BDT training gives preference to \Bp candidates that are isolated from the rest of the event.

Since no PID information is used in the BDT classifier, the efficiency for $\Bp \to D \Kp$ and $\Bp \to D \pip$ decays is similar, with insignificant variations arising from small differences in the decay kinematics. The selection requirement applied to the BDT response is optimised by minimising the relative statistical uncertainty on the $R_{SS/OS}$ observable, as measured using the fit described in Sec.~\ref{sec:mass_fit}. PID information from the RICH detectors is used to improve the purity of the $\Bp \to D\Kp$ samples. A strict PID requirement is applied to the companion kaon in $\Bp \to D \Kp$ to suppress contamination from $\Bp \to D \pip$ decays where the companion pion is misidentified as a kaon. The requirement is around 70\% efficient, and genuine $\Bp \to D \Kp$ decays failing the requirement are placed into the $\Bp \to D \pip$ sample. Less than 0.5\% of genuine $\Bp \to D \pip$ decays pass the kaon PID requirement, and are placed into the $\Bp \to D \Kp$ sample. This results in a $\Bp \to D \pip$  background of around 5\% relative to the correctly identified $\Bp \to D \Kp$ signal. Background from the $\Bp \to[\KS\pip\pim]_D h^+$ decay, which has a branching fraction around ten times larger than the signal, is suppressed by placing PID requirements on both the kaon and pion produced in the $D$ decay. 

For long \KS candidates, the square of the flight distance significance with respect to the PV is required to be greater than 100 to suppress background from $\Bp \to [\Kp\pim\pip\pim]_D h^+$ decays. Background from charmless $B$ decays such as $B^+ \to K_s^0 \Km \Kp \pip$, which peaks at the same invariant mass as the signal, is suppressed by requiring that the flight distance of the $D$ candidate divided by its uncertainty is greater than 2. Where multiple candidates are found in the same event, one candidate is chosen at random, leading to a reduction in the sample size of approximately 2\%.

For several quantities used in the selection and analysis of the data,
a kinematic fit~\cite{Hulsbergen:2005pu} is imposed on the full \Bp decay chain. 
Depending on the quantity being calculated, 
the \D and \KS candidates may be constrained to have their known masses~\cite{PDG2018}.
The fit also constrains the \Bp candidate momentum vector to point towards the associated PV, defined as the PV for which the candidate has the smallest $\chi_{\text{IP}}^2$.
These constraints improve the resolution of the calculated quantities, and thus help enhance the separation between signal and background decays.
Furthermore, they improve the mass-squared resolution, which is important for identifying the Dalitz region assignment.

\begin{figure}
\centering
\includegraphics[width=0.42\textwidth]{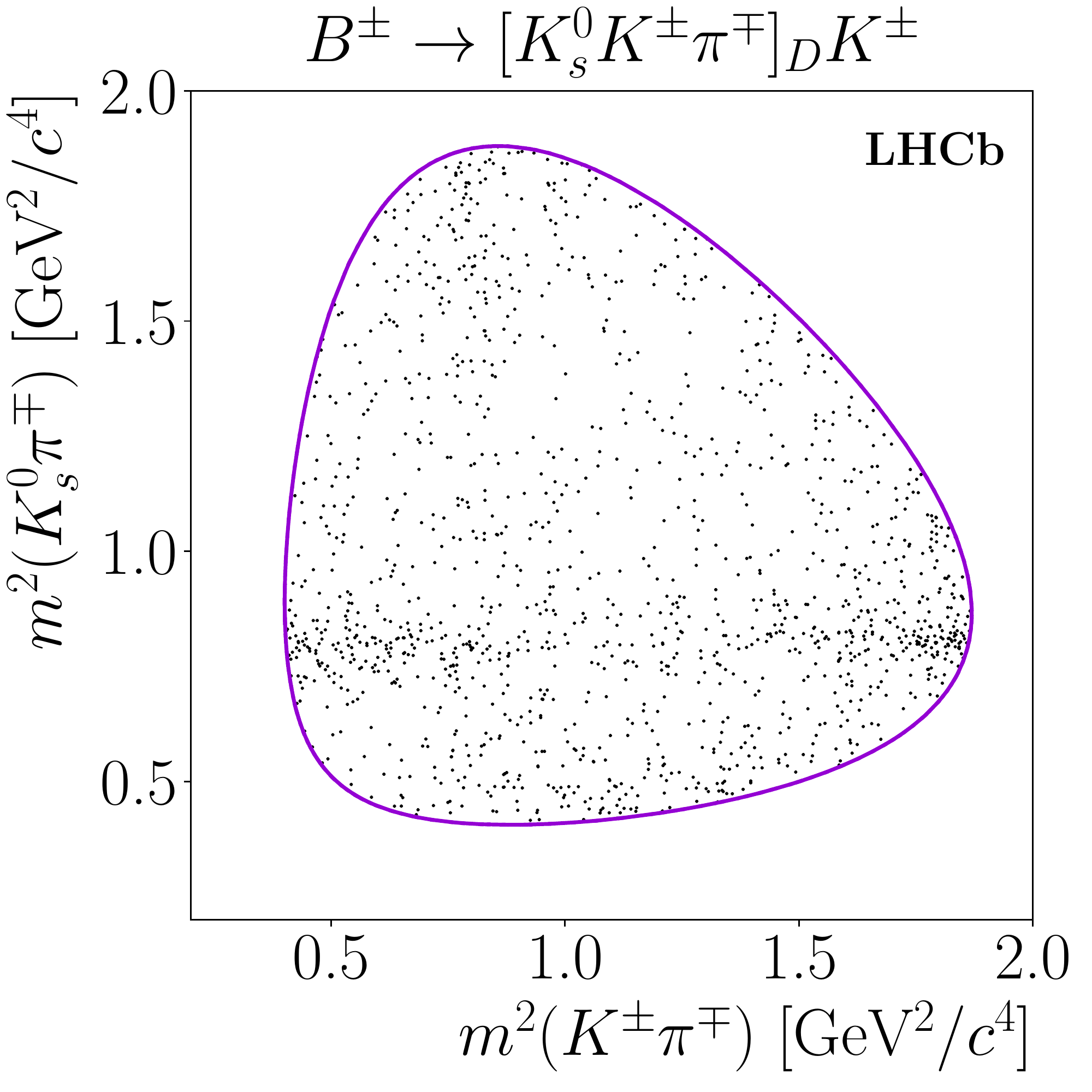}
\includegraphics[width=0.42\textwidth]{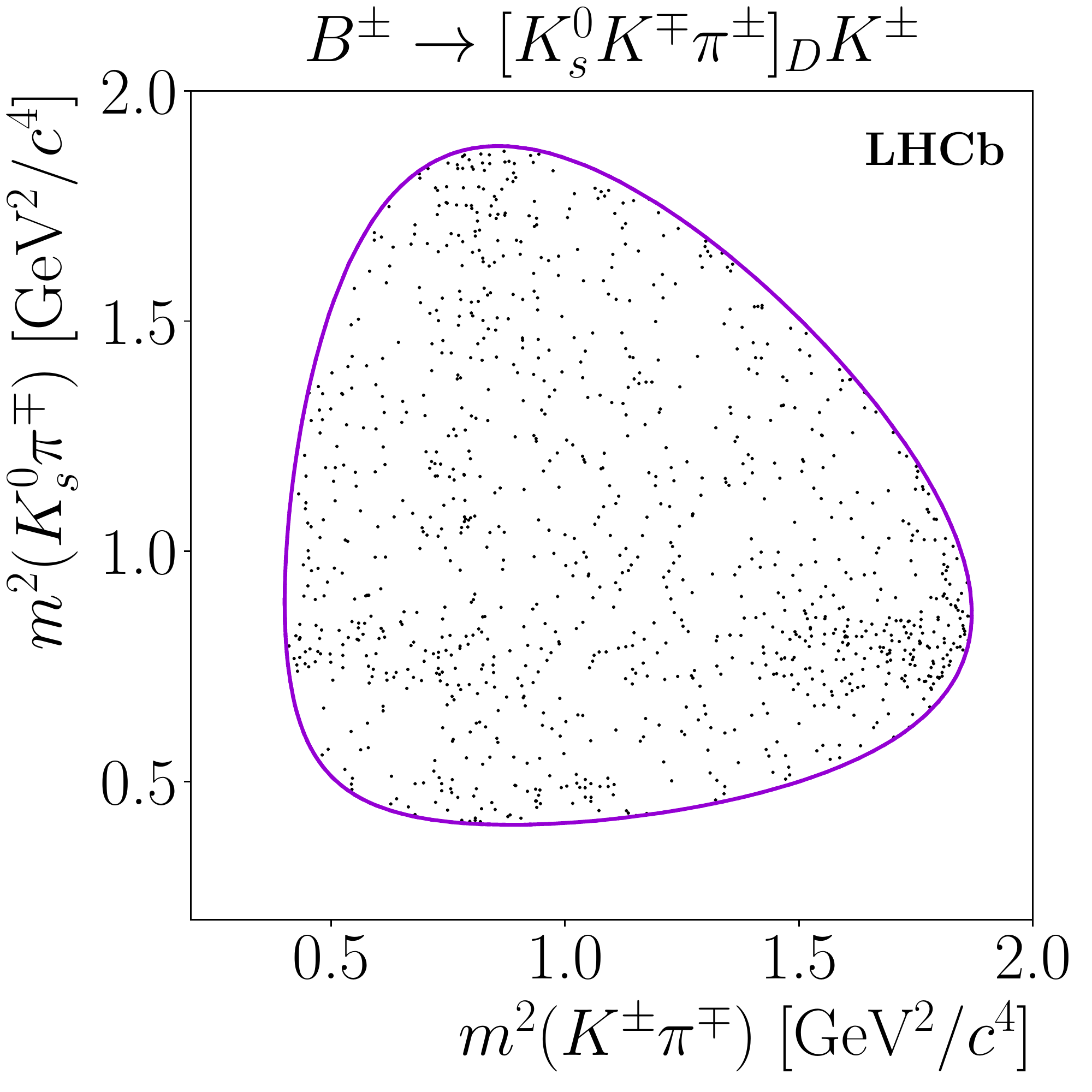}\\ \vspace{0.5cm}
\includegraphics[width=0.42\textwidth]{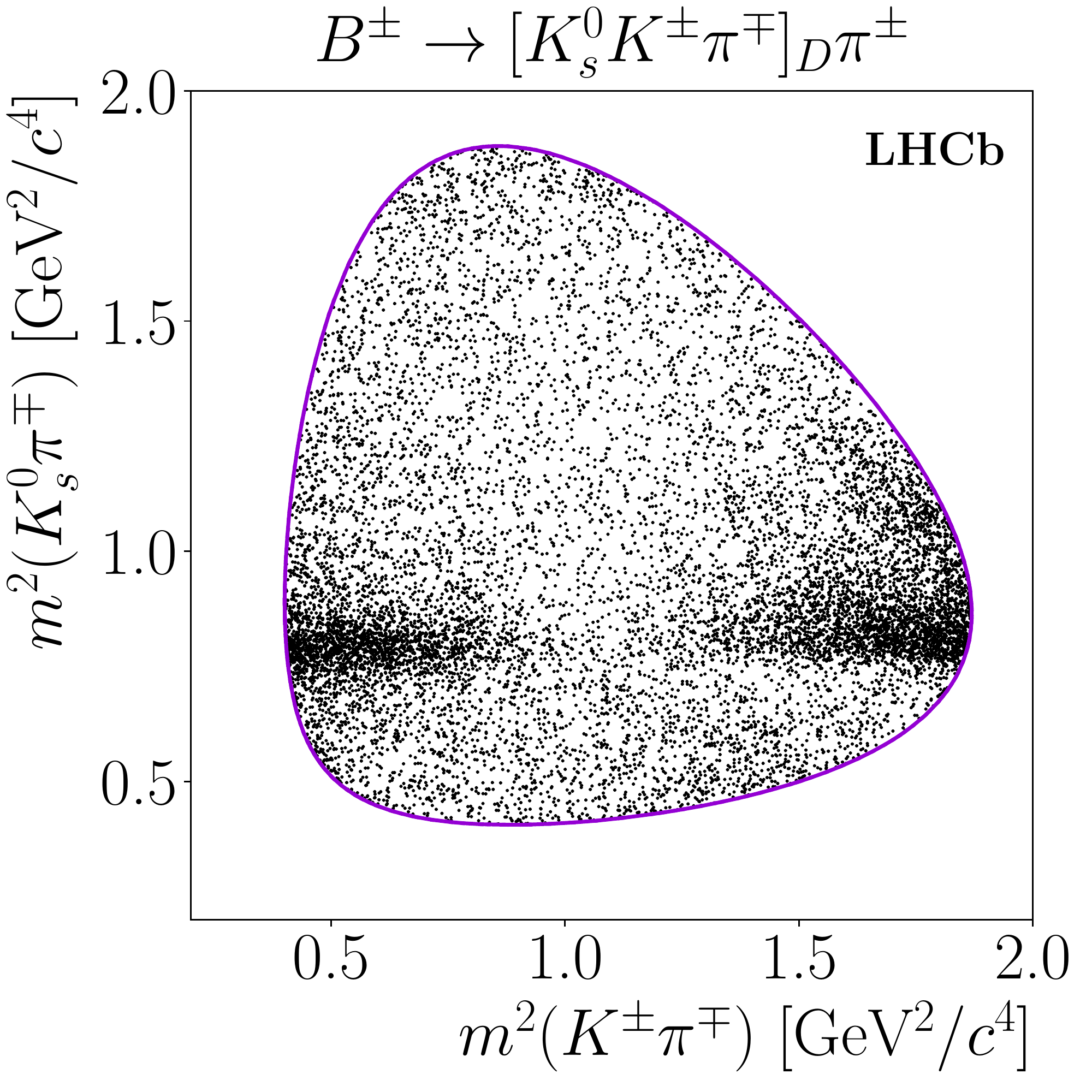}
\includegraphics[width=0.42\textwidth]{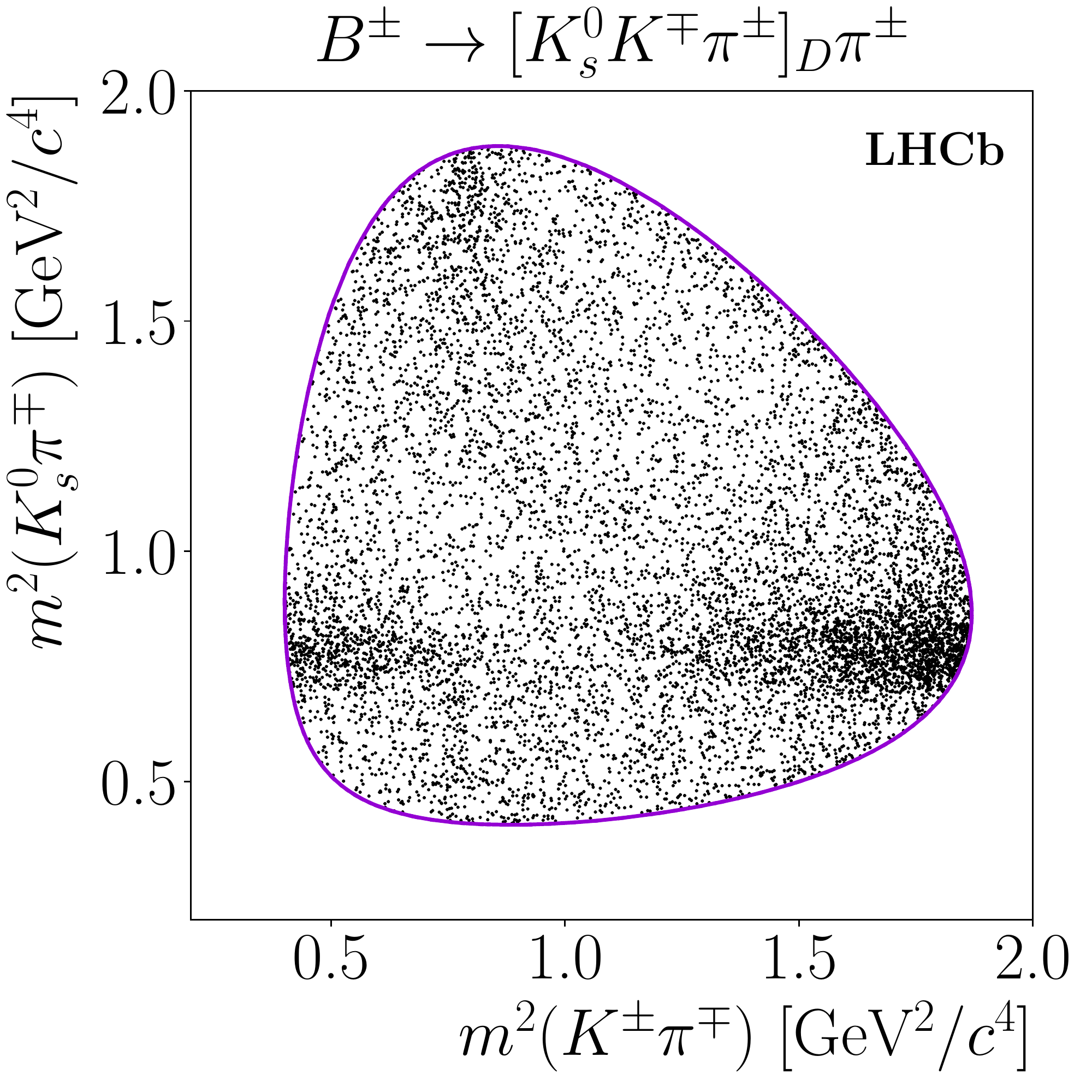}
\caption{\small $D\to \KS\Kp\pim$ Dalitz plots from the SS (left) and OS (right) data samples, for $\Bp \to D \Kp$ (top) and $\Bp \to D \pip$ (bottom). Both long and downstream $\KS$ decays are included. The purple lines indicate the kinematic boundary.}
\label{fig:dalitz}
\end{figure}

The Dalitz plots for selected candidates in the signal region $\pm 25\mevcc$ around the \Bp mass are shown in 
Fig.~\ref{fig:dalitz}; within this region, background from decays involving no charm meson constitute less than 5\% of the total sample. The Dalitz coordinates are calculated from the kinematic fit with all mass constraints applied.
A band corresponding to the intermediate state, $D\to\Kstar(892)^-\Kp$ is visible in each plot.

\section{Invariant-mass fit}
\label{sec:mass_fit}

In order to measure the \CP observables introduced in Sec.~\ref{sec:formalism}, an extended binned maximum-likelihood fit to the invariant-mass distributions of the $B$ meson candidates in the range between 5080\mevcc and 5700\mevcc is performed. The fit is performed simultaneously to all decay categories, in order to enable sharing of common parameters. A total of 16 categories are included in the fit: ($DK$, $D\pi$) $\times$ (SS, OS) $\times$ (long, downstream) $\times$ ($\Bp$, $\Bm$). The fit range is between 5080\mevcc and 5700\mevcc in the \Bpm candidate invariant mass. The fit is performed separately for candidates within the \Kstarp region and those outside.

To model the invariant-mass distribution, a total fit probability density function (PDF) is created from several signal and background components. Most of these are modelled using simulated signal and background samples reconstructed as the signal decay and passing all selection requirements. The components are:
\begin{enumerate}
    \item Signal $\Bp \to D \Kp$ and $\Bp \to D \pip$ decays, described by the sum of two Crystal Ball functions~\cite{CrystalBall} with a freely varying common mean and width, and tail parameters fixed from simulation. A single freely varying parameter relates all $\Bp \to D \Kp$ widths to their $\Bp \to D \pip$ counterparts. SS and OS decays share all shape parameters, but long and downstream decays have separate freely varying widths due to the differences in invariant-mass resolution. All shape parameters are identical for $\Bp$ and $\Bm$ decays. 
    
    \item Combinatorial background, described by an exponential function with a freely varying exponent in each (SS, OS) $\times$ (long, downstream) category. The combinatorial background yield freely varies in each ($DK$, $D\pi$) $\times$ (SS, OS) $\times$ (long, downstream) category, but is required to be the same in \Bp and \Bm.
    
    \item Partially reconstructed background from the $\Bp \to (\Dstarz \to D \{\piz/\gamma\}) h^+$, \mbox{$\Bp \to D h^+ \{\piz\}$}, and \mbox{$\Bz \to (D^{*-} \to D \{\pim\}) h^+$} decays, where the particle in braces is not reconstructed. These components sit at lower invariant-mass values than the signal, and are described by PDFs constructed from a parabolic function to describe the decay kinematics. This function is convolved with the sum of two Gaussian functions with a common mean in order to describe the detector resolution, as further described in Ref.~\cite{LHCb-PAPER-2017-021}. All shape parameters are fixed from simulation.
    All partially reconstructed background yields vary freely, but the $\Bz \to D^{*-} h^+$ component yields  are required to be equal in the $B^-$ and $B^+$ samples; the fast $\Bz$ oscillation renders \CP violation effects negligible in this time-integrated measurement.
    
    \item Partially reconstructed background from $B_s^0 \to D \Kp \{\pim\}$ decays, where the pion produced in the $B_s^0$ decay is not reconstructed, contributes in the $\Bp \to D \Kp$ samples. These decays are modelled using a PDF with fixed shape parameters based on the $m(DK)$ distribution observed in Ref.~\cite{LHCb-PAPER-2014-036}.
    The yield of this component freely varies in the SS and OS samples, but the yields  are required to be equal in the $B^-$ and $B^+$ samples as the fast $\B_s^0$ oscillation renders \CP violation effects negligible.
    
    \item Charmless background contributions remain in the $\Bp \to D \Kp$ samples after application of the \D-meson flight requirement described in Sec.~\ref{sec:selection}. They are estimated using fits to the $B^+$ candidate invariant-mass distributions in data, where candidates falling in the lower sidebands of the \D candidate invariant-mass spectra are considered. The charmless contributions are included as fixed-shape Gaussian functions from simulation, with fixed yields as determined by the sideband fits.
    
    \item Backgrounds from particle misidentification, which arise due to the imperfect efficiency of RICH PID requirements applied to companion hadrons in order to separate $\Bp \to D \Kp$ and $\Bp \to D \pip$ decays. The efficiencies of the PID requirements are determined using calibration samples of high-purity decays that can be identified without the use of RICH information~\cite{LHCb-DP-2018-001}. Given a PID efficiency $\epsilon_{\text{PID}}^{K} \sim 0.7$ for $\Bp \to D \Kp$ decays, a fixed fraction $(1-\epsilon_{\text{PID}}^{K})$ of the total $\Bp \to D \Kp$ signal yield is assigned to a PDF in the corresponding $\Bp \to D \pip$ sample. This component is described by a Crystal Ball function with all shape parameters fixed to those found in simulation; due to the companion hadron misidentification, this component falls below the nominal \Bp mass. In the same fashion, a small component is included in the $\Bp \to D \Kp$ sample to model misidentified $\Bp \to D \pip$ decays, with a yield that freely varies to $\sim 0.4\%$ of the total $\Bp \to D \pip$ yield. This component is also described by a Crystal Ball function, with all shape parameters fixed to the values found in simulation.
\end{enumerate}

In order to measure \CP asymmetries, the detection asymmetries for \Kpm and \pipm mesons must be taken into account. In the fit, a detection asymmetry of $(-0.51 \pm 0.28)$\% is assigned for each kaon in the final state, primarily due to the fact that the nuclear interaction length of \Km mesons is shorter than that of \Kp mesons. 
The value used is computed by comparing the charge asymmetries in $\Dp \to \Km \pip\pip$ and $\Dp \to \KS \pip$ calibration samples, weighted to match the kinematics of the signal kaons~\cite{LHCb-PUB-2018-004}. 
The equivalent asymmetry for pions is smaller, $(-0.06 \pm 0.04)$\%~\cite{LHCb-PAPER-2016-054}. All measured \CP asymmetries are also corrected in the fit for the asymmetry in \Bpm production, which has a value $(+0.14 \pm 0.07)\%$ based on measurements of this quantity made in Refs.~\cite{LHCb-PAPER-2017-021} and~\cite{LHCb-PAPER-2016-054}.

To measure the $R_{SS}^{DK/D\pi}$ and $R_{OS}^{DK/D\pi}$ observables, the raw signal yields are corrected for small differences in the total efficiency for selecting $\Bp \to D \Kp$ and $\Bp \to D \pip$ decays. The efficiency ratio is found to be $\frac{\epsilon(DK)}{\epsilon(D\pi)} = 1.012 \pm  0.016$, which is employed as a fixed correction term in the fit. A similar correction is applied to the $R_{SS/OS}$ observable, to account for differences in selection efficiency for SS and OS decays caused by efficiency variation across the Dalitz plot. The correction is determined using simulated $\Bp \to D \pip$ decays and the $D \to \KS \Kp \pim$ and $D \to \KS \Km \pip$ amplitudes measured by LHCb in Ref.~\cite{LHCb-PAPER-2015-026}. The correction is determined in bins across the Dalitz plot, and an average value is calculated to be $\eta = 1.090 \pm 0.008$ ($1.007 \pm 0.013$) within (outside) the \Kstarp region.

\begingroup
\renewcommand*{\arraystretch}{1.2}
\begin{table}[!t]
\centering
\caption{Signal yields summed over charge, as measured in each Dalitz region.}
\begin{tabular}{l l l}
\hline
 & non-\Kstarp region & \Kstarp region \\ \hline
$N_{\rm{SS}}^{D\Kpm}$ &  $\phantom{0}266 \pm 27$ &  $\phantom{0}715 \pm \phantom{0}37$ \\
$N_{\rm{OS}}^{D\Kpm}$ &  $\phantom{0}336 \pm 27$ &  $\phantom{0}217 \pm \phantom{0}22$ \\
$N_{\rm{SS}}^{D\pipm}$ & $3304 \pm 73$ & $8977 \pm 106$ \\
$N_{\rm{OS}}^{D\pipm}$ & $4686 \pm 76$ & $3471 \pm  \phantom{0}66$ \\  \hline
\end{tabular}
\label{tab:yields}
\end{table}
\endgroup

\begin{figure}[!t]
  \begin{center}
   \includegraphics[width=0.49\linewidth]{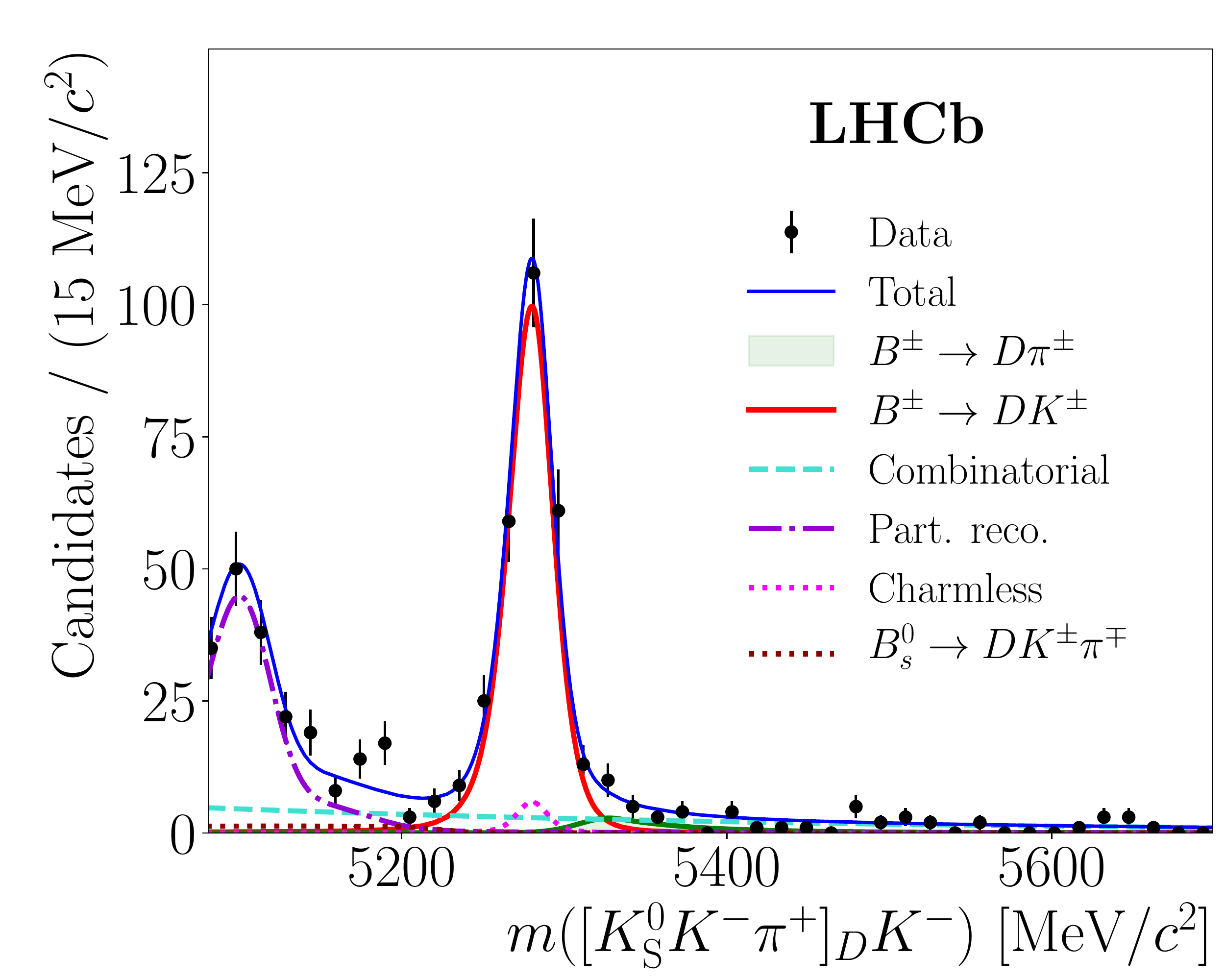}
   \includegraphics[width=0.49\linewidth]{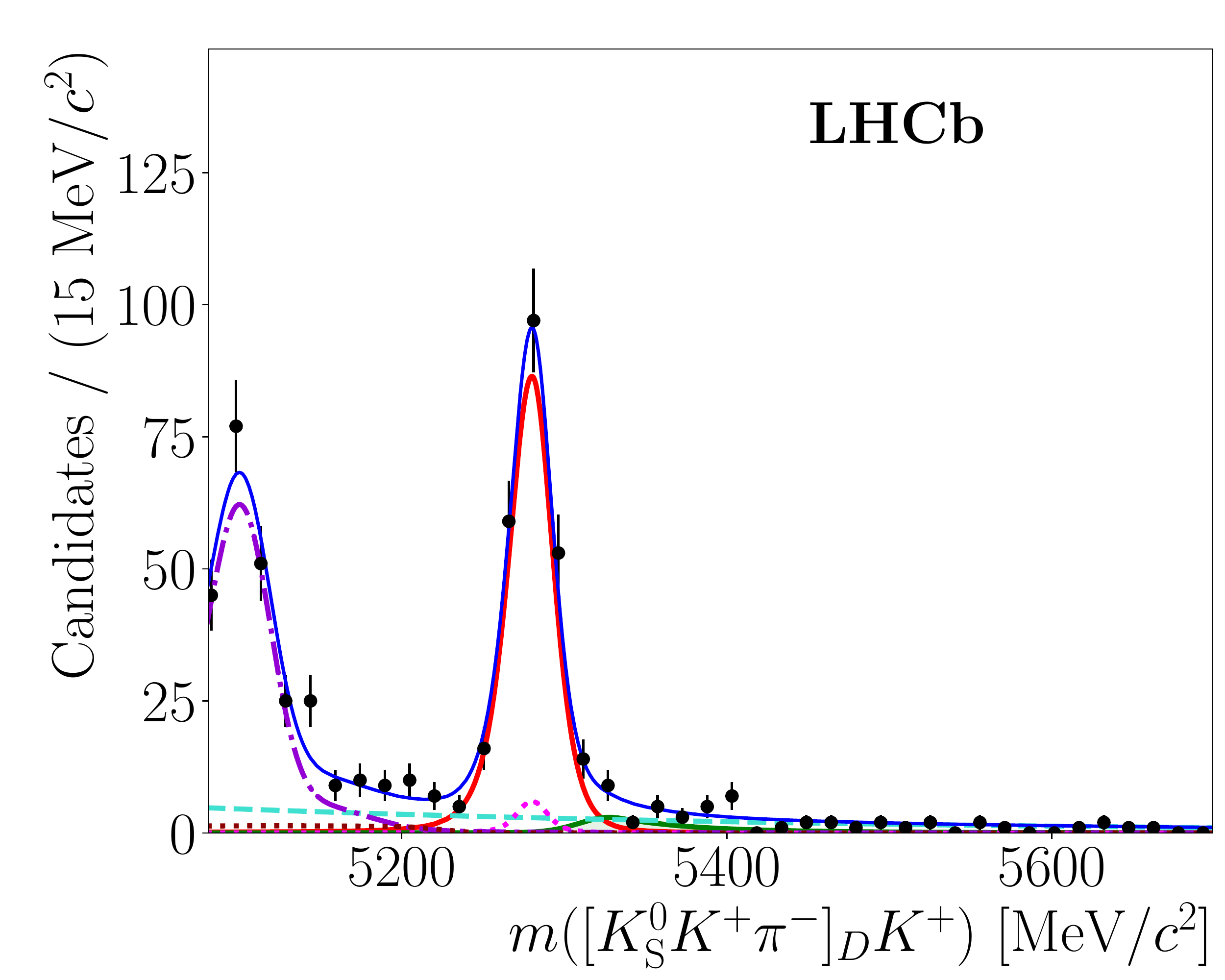}\\
   \includegraphics[width=0.49\linewidth]{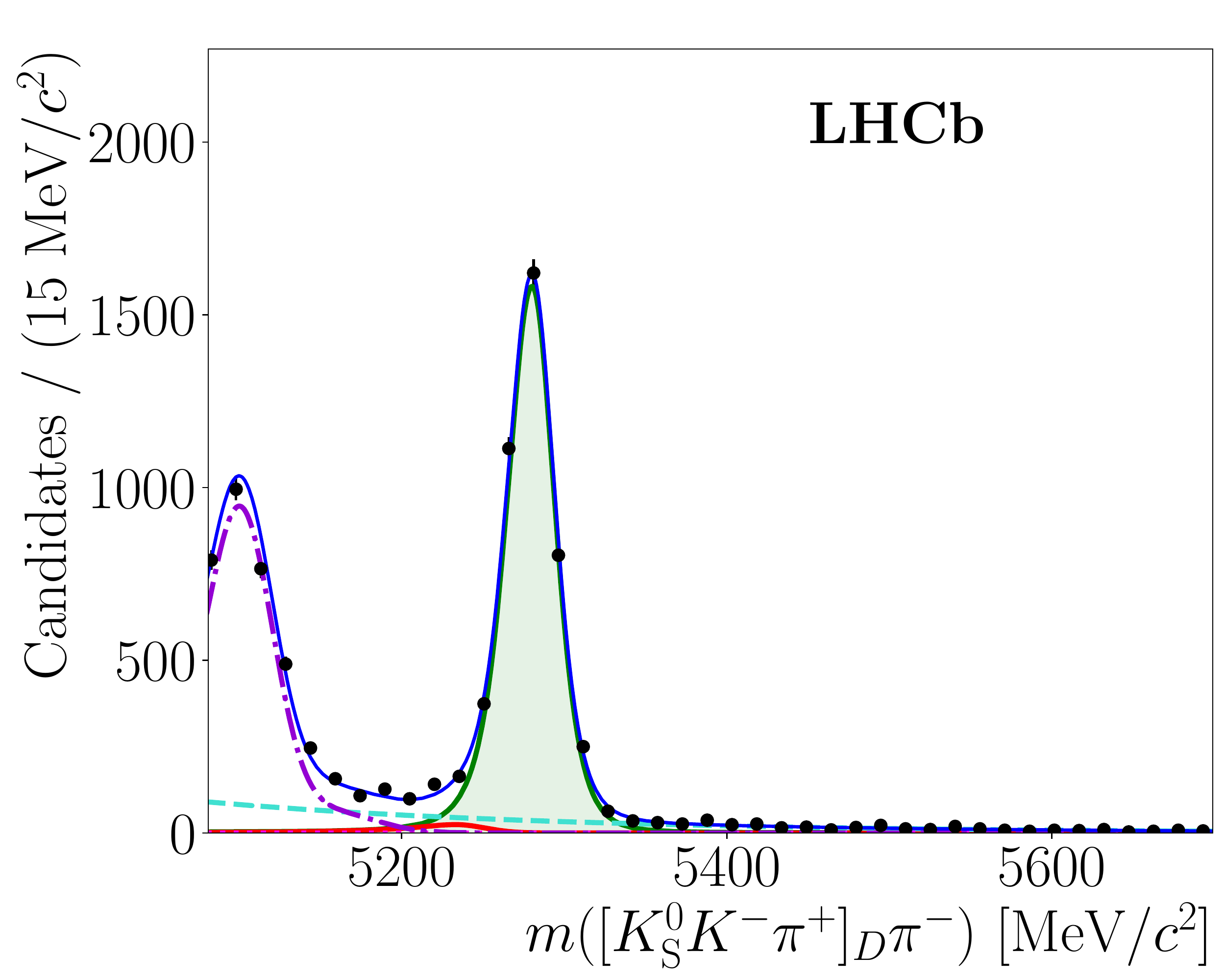}
   \includegraphics[width=0.49\linewidth]{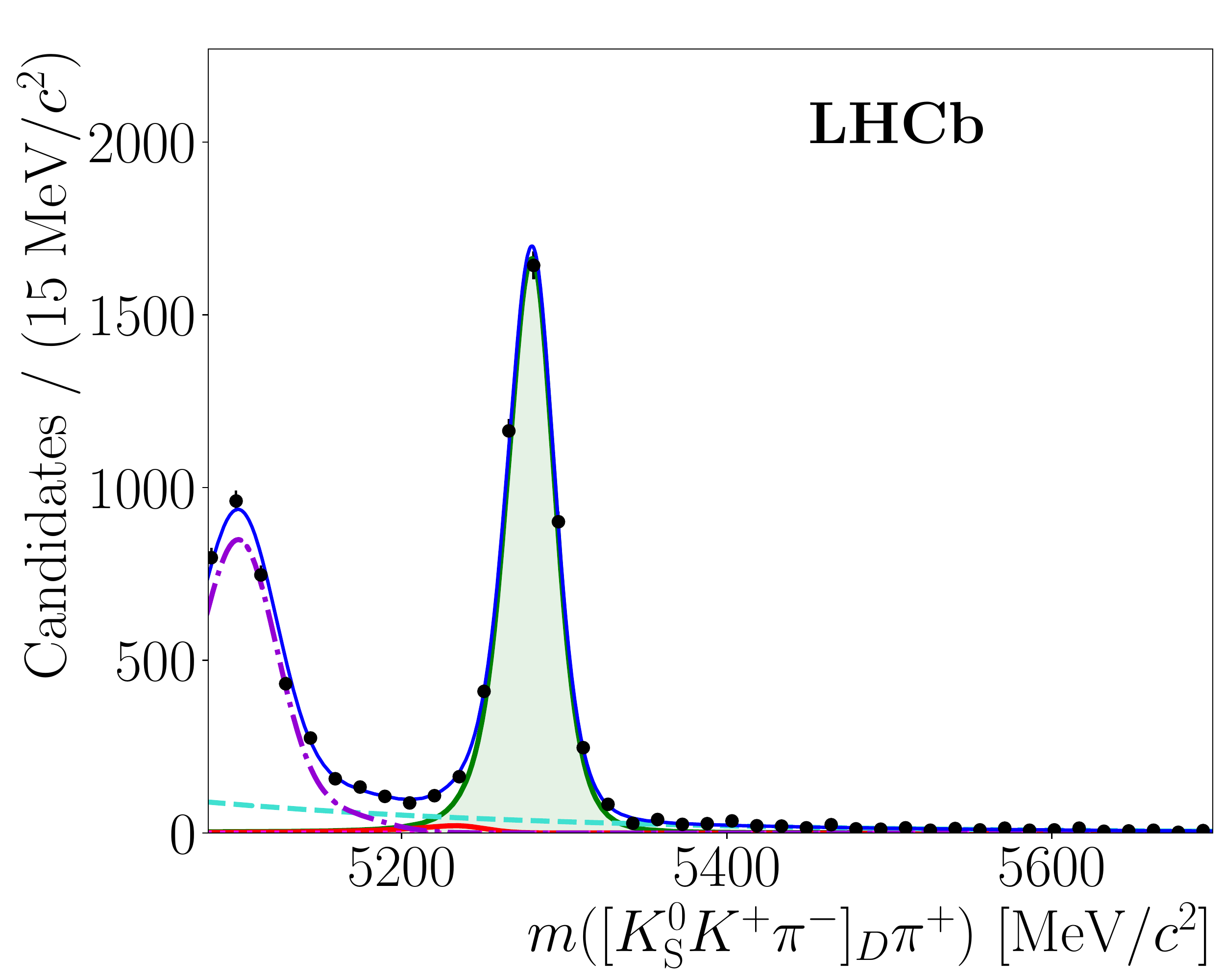}\\
   \end{center}
   \caption{Invariant mass of SS $\Bpm \to [\KS \Kpm \pimp]_D h^{\pm}$ candidates within the \Kstarp region; candidates containing both long and downstream $K_s^0$ mesons are shown.
  \label{fig:KsKPi_Kst}}
\end{figure}

\begin{figure}[!t]
  \begin{center}
 \includegraphics[width=0.49\linewidth]{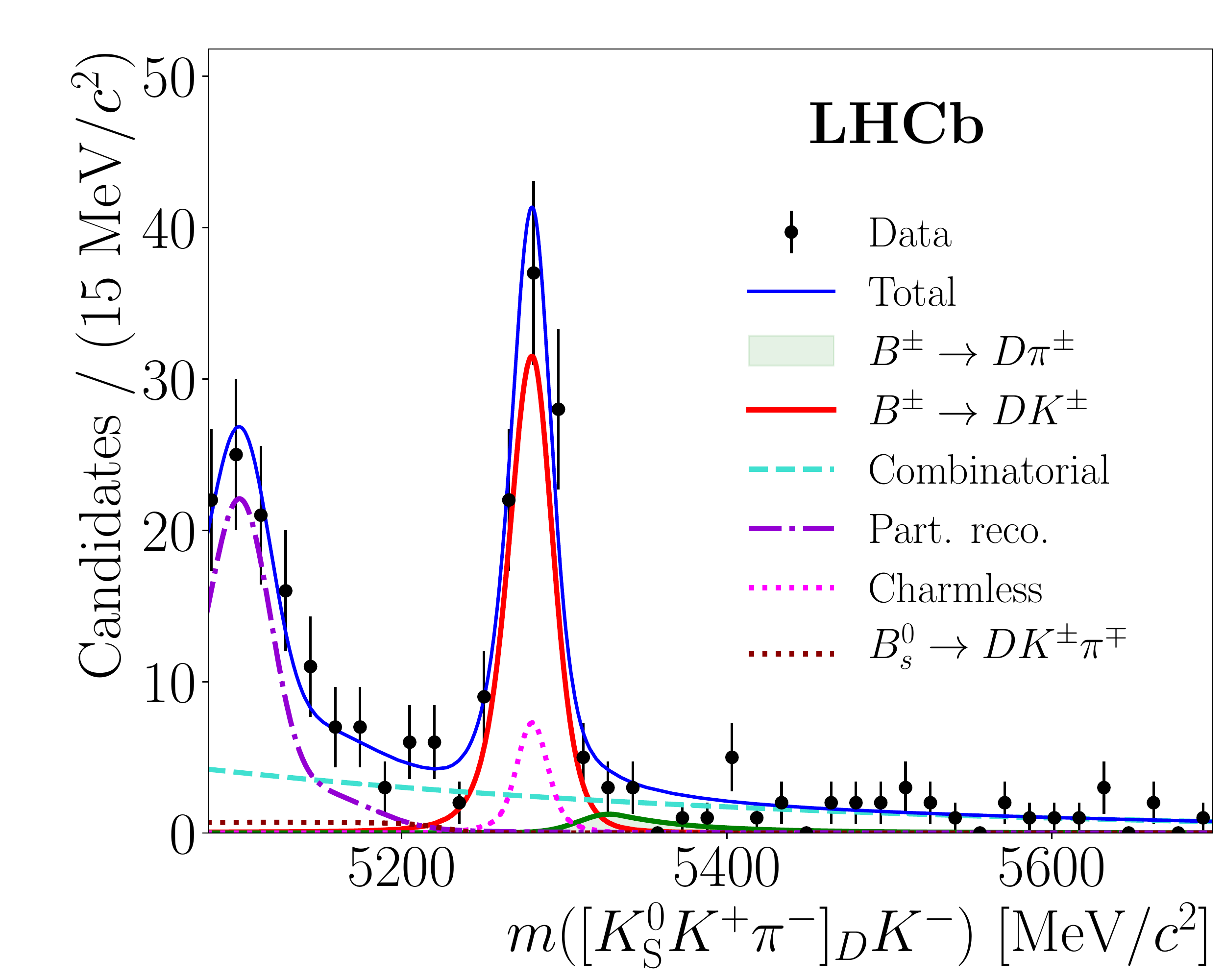}
   \includegraphics[width=0.49\linewidth]{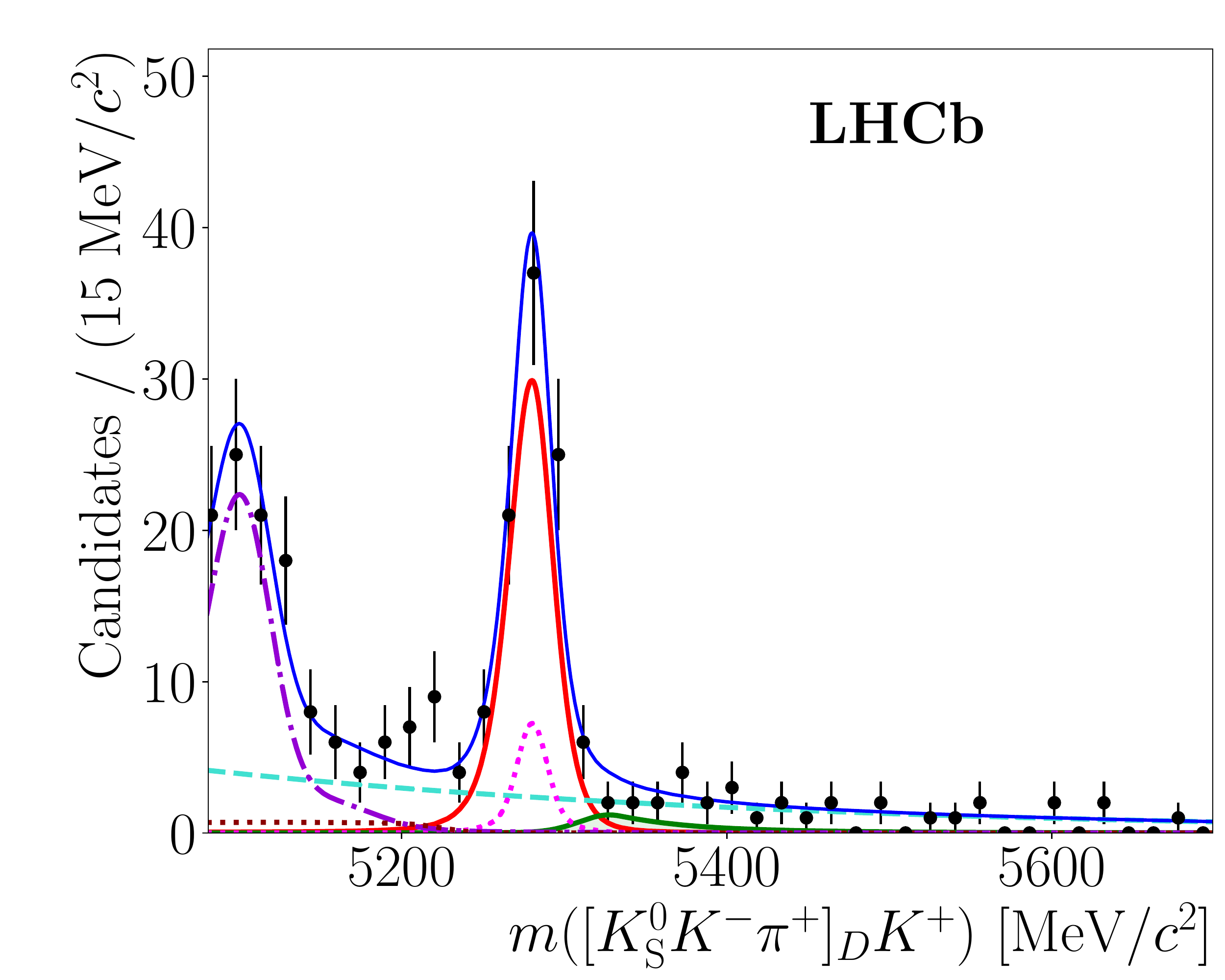}\\
   \includegraphics[width=0.49\linewidth]{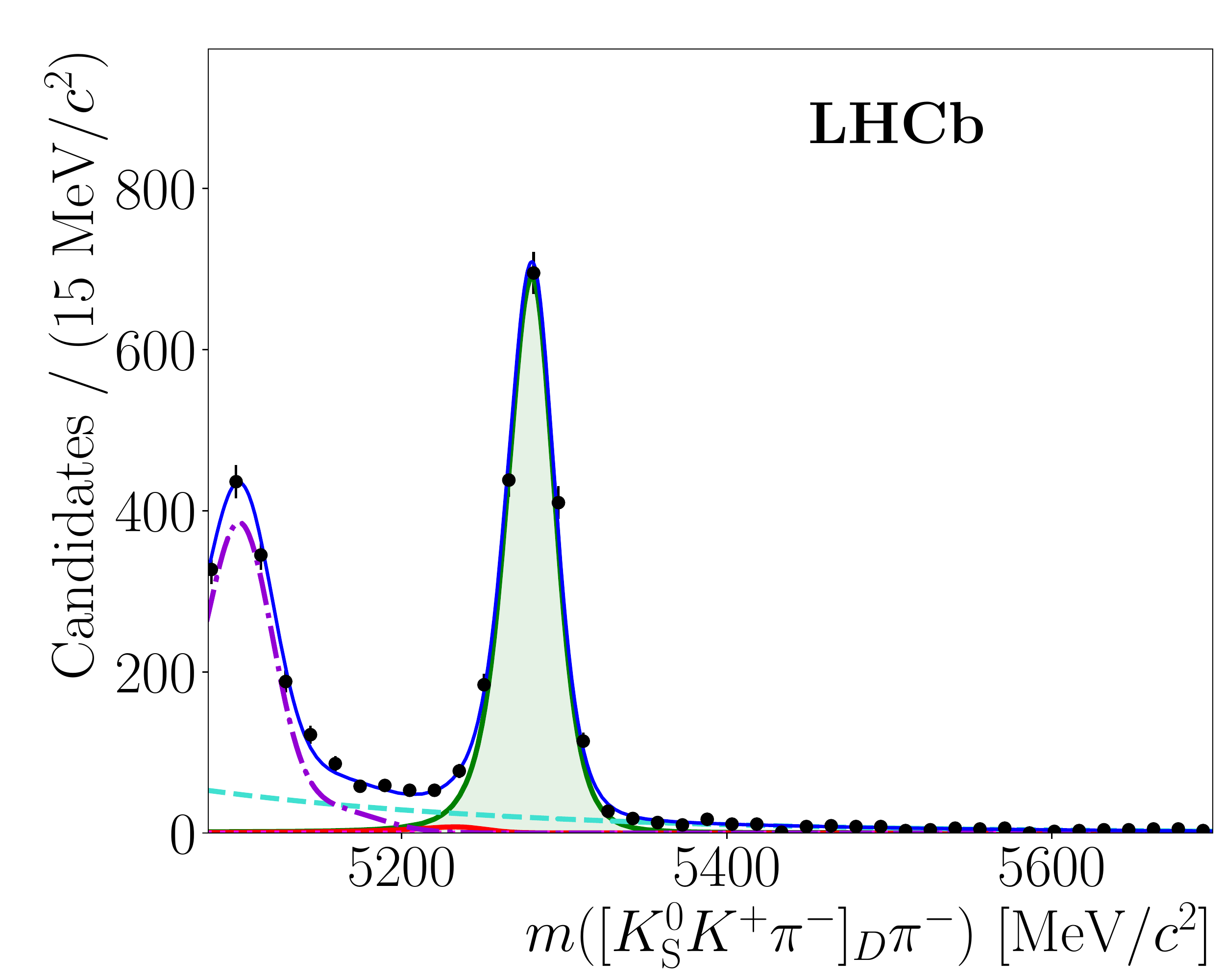}
   \includegraphics[width=0.49\linewidth]{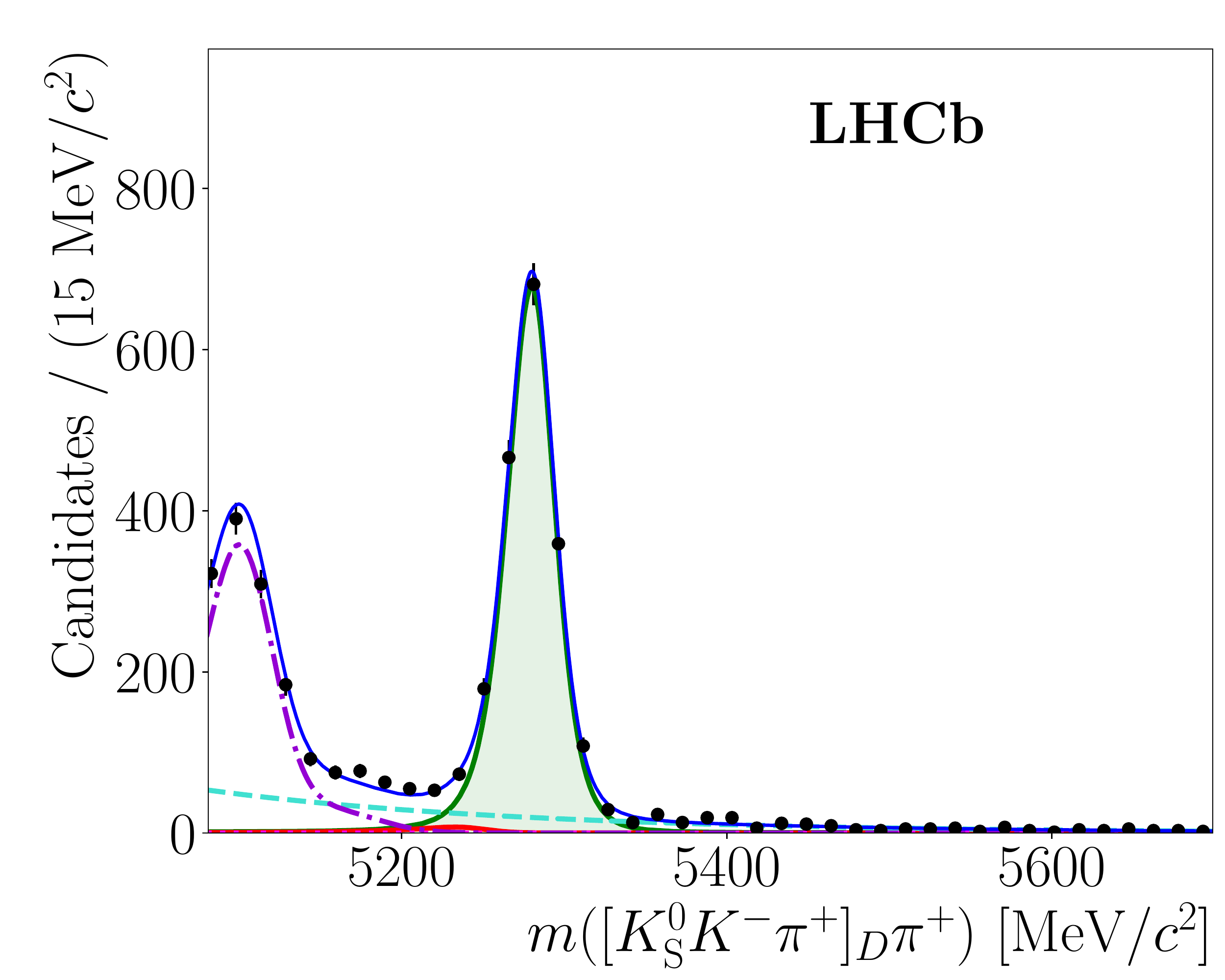}\\
  \end{center}
  \caption{Invariant mass of OS $\Bpm \to [\KS \Kmp \pipm]_D h^{\pm}$ candidates within the \Kstarp region. The fit components are detailed in the legend of Fig.~\ref{fig:KsKPi_Kst}.
  \label{fig:KsPiK_Kst}}
\end{figure}

\begin{figure}[!t]
  \begin{center}
  \includegraphics[width=0.49\linewidth]{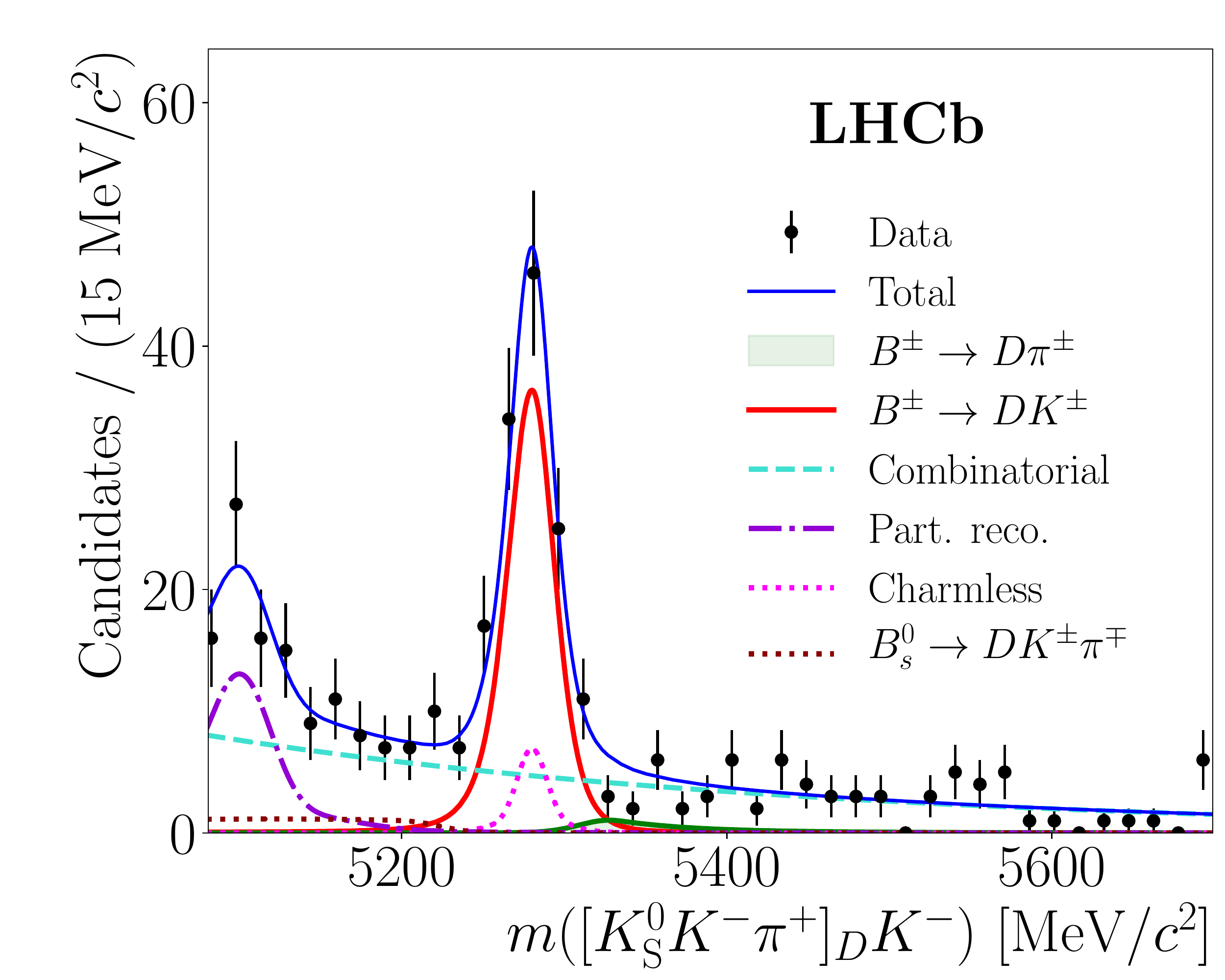}
   \includegraphics[width=0.49\linewidth]{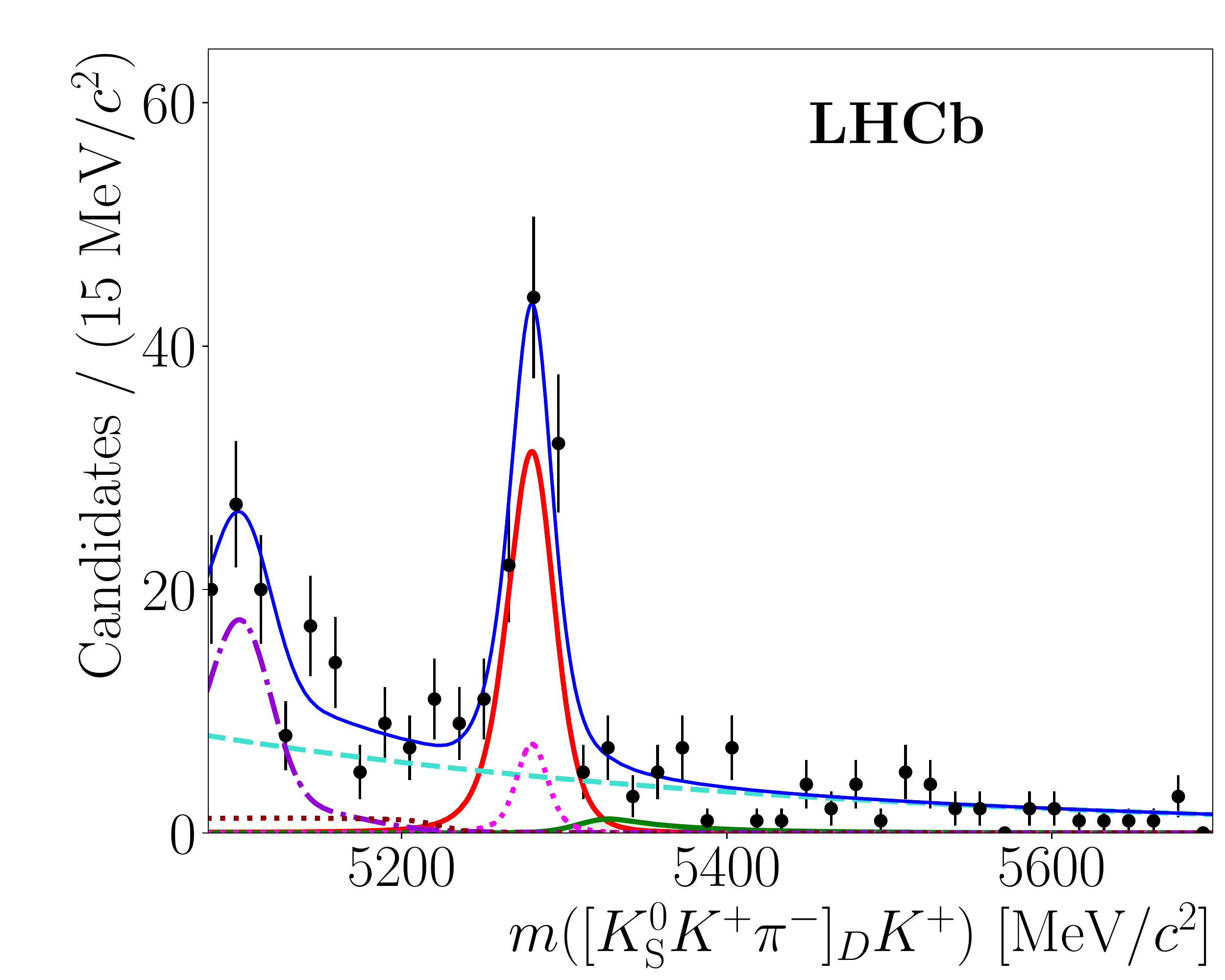}\\
   \includegraphics[width=0.49\linewidth]{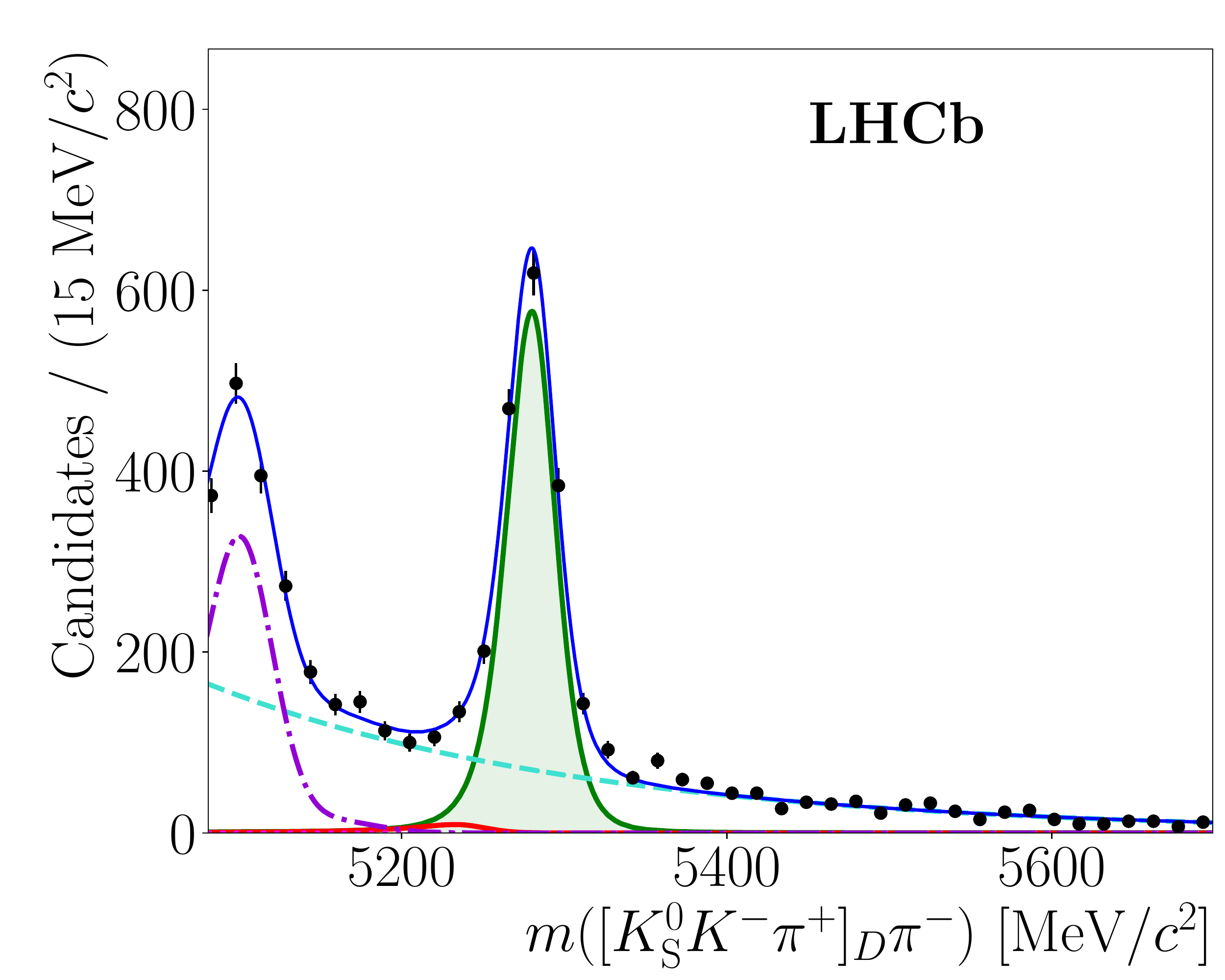}
   \includegraphics[width=0.49\linewidth]{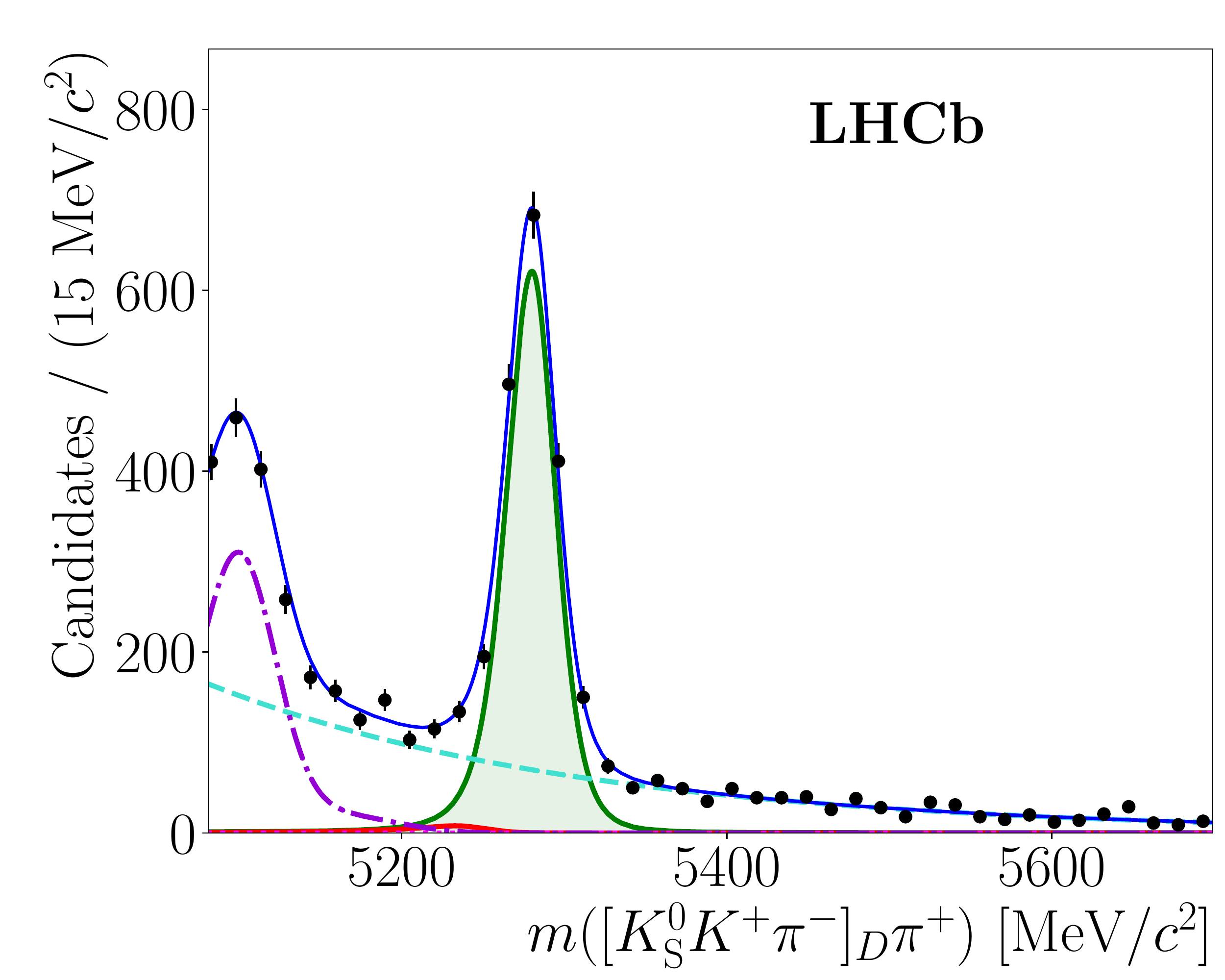}\\
       \end{center}
  \caption{Invariant mass of SS $\Bpm \to [\KS \Kpm \pimp]_D h^{\pm}$ candidates in the non-\Kstarp region. The fit components are detailed in the legend of Fig.~\ref{fig:KsKPi_Kst}.
  \label{fig:KsKPi_full}}
\end{figure}

\begin{figure}[!t]
  \begin{center}
   \includegraphics[width=0.49\linewidth]{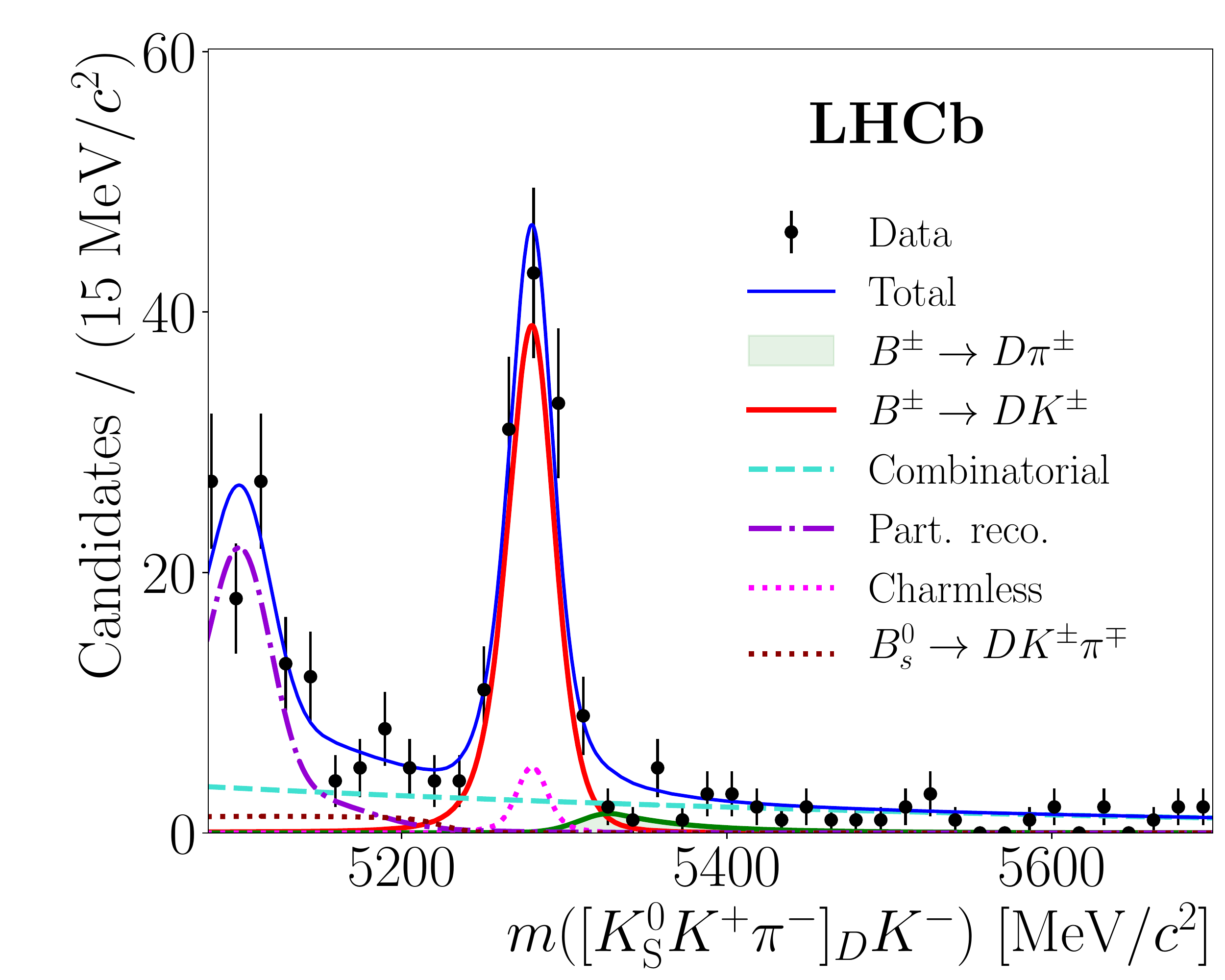}
   \includegraphics[width=0.49\linewidth]{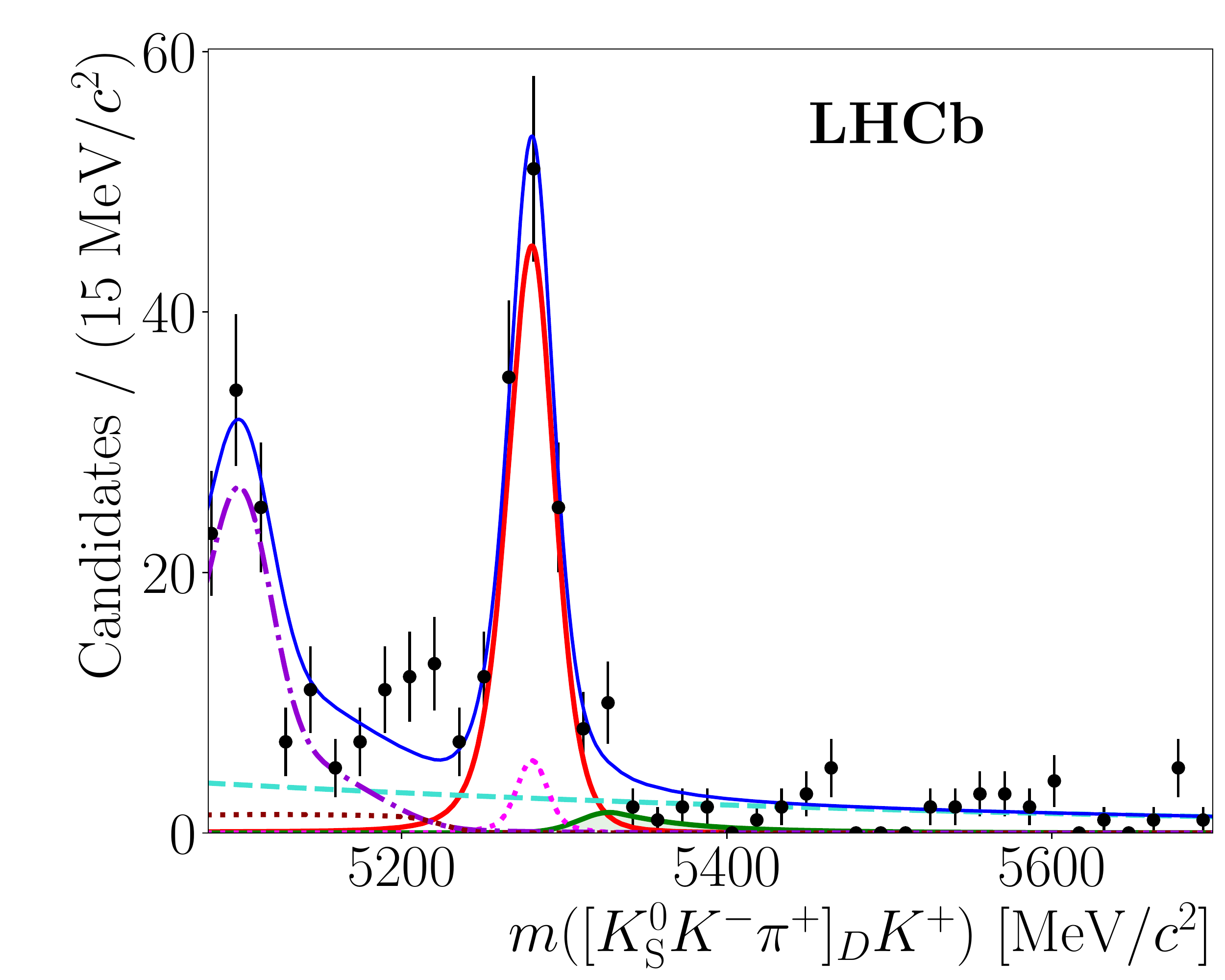}\\
   \includegraphics[width=0.49\linewidth]{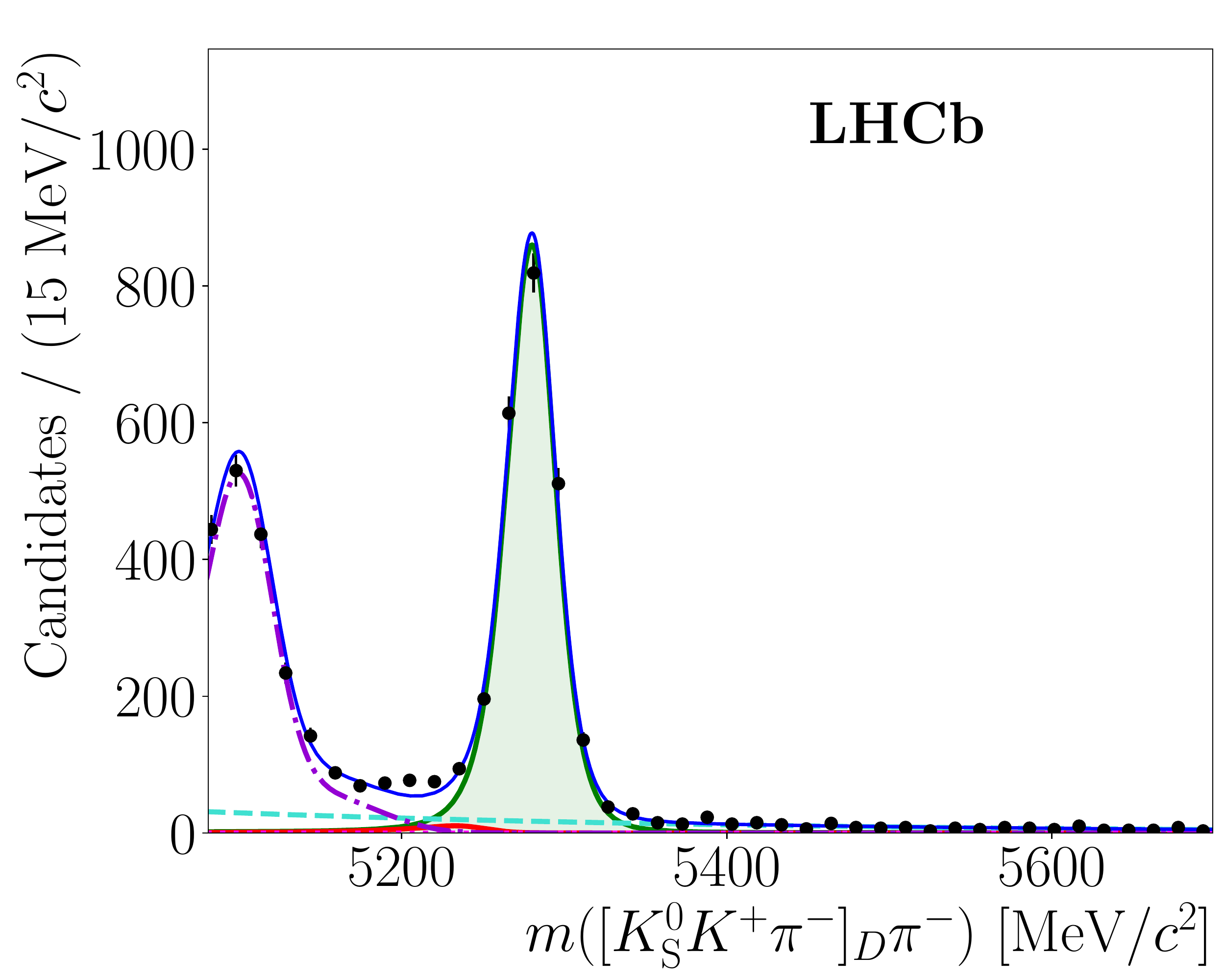}
   \includegraphics[width=0.49\linewidth]{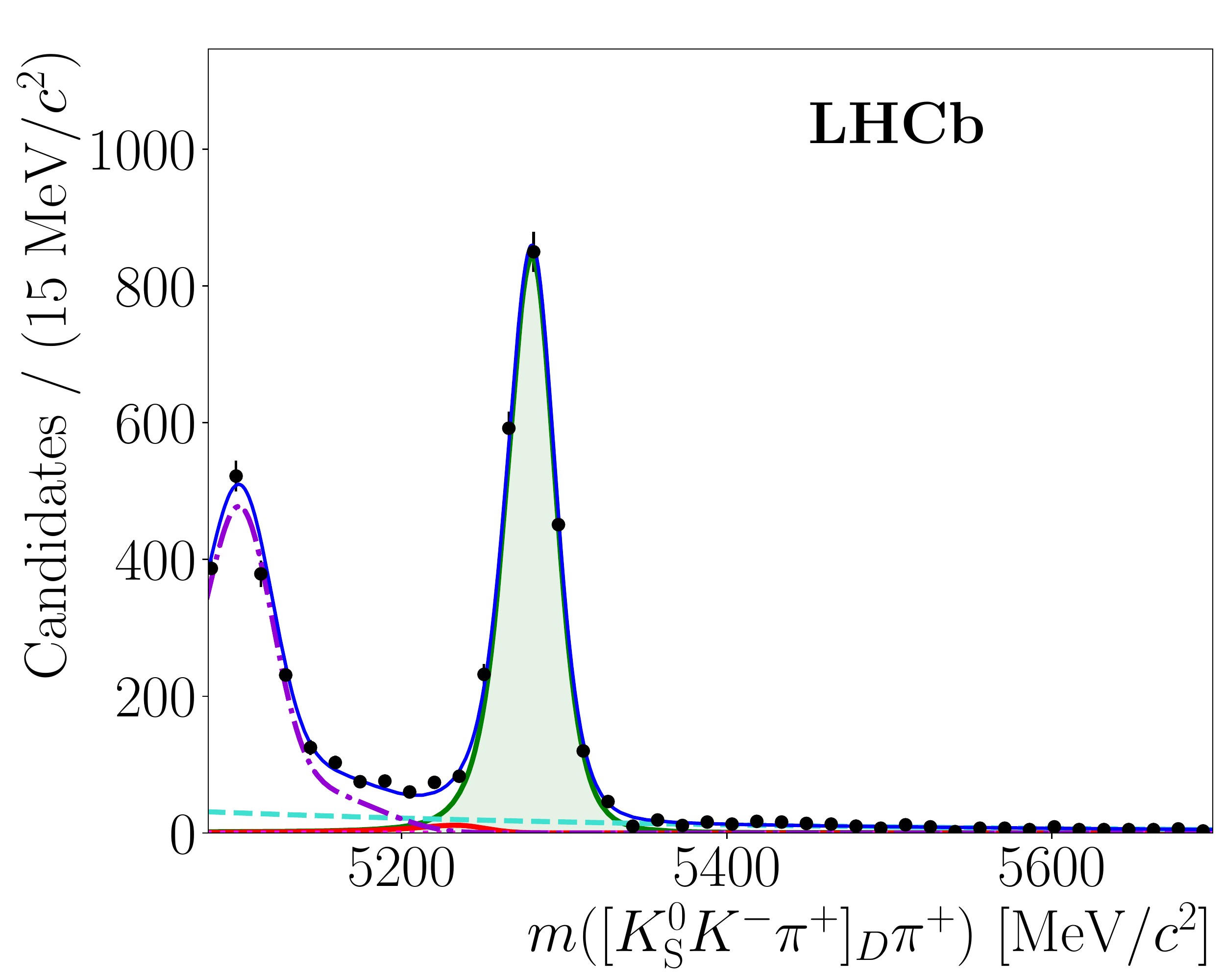}\\
    \end{center}
  \caption{Invariant mass of OS $\Bpm \to [\KS \Kmp \pipm]_D h^{\pm}$ candidates in the non-\Kstarp region. The fit components are detailed in the legend of Fig.~\ref{fig:KsKPi_Kst}.
  \label{fig:KsPiK_full}}
\end{figure}

Figs.~\ref{fig:KsKPi_Kst}--\ref{fig:KsPiK_full} show the $B$ meson invariant-mass distributions for all selected candidates, with the results of the fit overlaid; the long and downstream $K_s^0$ categories are shown together. In Tab.~\ref{tab:yields}, the measured signal yields for each $D$ final state are provided for both the \Kstarp and the non-\Kstarp regions. The fit strategy is validated using pseudoexperiments, and is found to be unbiased for all parameters.


\section{Systematic uncertainties}
\label{sec:systematics}

All of the \CP observables measured in this work are constructed as ratios of topologically identical final states. As such, the majority of potential systematic uncertainties cancel with the residual systematic uncertainties detailed here. Small differences in efficiency between $\Bp \to D \Kp$ and $\Bp \to D \pip$ decays are corrected using simulation as described in Sec.~\ref{sec:mass_fit}, where the uncertainty on the correction arises due to the finite size of the simulated samples. The correction is varied within its uncertainty to determine the systematic uncertainty. The variation in efficiency across the Dalitz plot causes a difference in the total efficiency of SS and OS decays. An appropriate correction is applied to the $R_{SS/OS}$ observable, with an uncertainty arising from the use of a binned procedure to calculate the average correction. 

Several fixed shape parameters are used in the fit, including the signal tail parameters and background PDFs. All fixed shape parameters are determined from fits to simulated samples, and are varied to calculate the propagated systematic uncertainty.
Charmless backgrounds are modelled as fixed yield components in the invariant-mass fit. The yields are varied within their respective uncertainties to determine the systematic uncertainty. Each charmless component has a fixed \CP asymmetry of zero in the fit; their asymmetries are independently varied according to a Gaussian of width 0.1 to determine the systematic uncertainty. This width chosen to align with the degree of \CP asymmetry observed in the charmless background present in measurements of $\Bp \to [h^+h^-]_D h^+$ decays.~\cite{LHCb-PAPER-2016-003,LHCb-PAPER-2017-021}.

All measured \CP asymmetries are corrected for the \Bpm production asymmetry as well as for the kaon and pion detection asymmetries where relevant. These corrections are applied as fixed terms in the invariant-mass fit, and are varied within their associated uncertainties to determine the systematic uncertainty. A fixed PID efficiency is used to determine the fraction of $\Bp \to D \Kp$ signal decays that are misidentified as $\Bp \to D \pip$. This efficiency is known within 1\% relative uncertainty, and is varied within this range to determine the systematic uncertainty.

The systematic uncertainties for each \CP observable, quoted as a percentage of the statistical uncertainty, are listed in Tabs.~\ref{tab:systematics_Kst_Y} and~\ref{tab:systematics_Kst_N}. The category \textit{Eff} relates to efficiency corrections, \textit{PDF} to fixed shape parameters, \textit{Cls} to charmless background yields and asymmetries, \textit{Asym} to asymmetry corrections, and \textit{PID} to the PID efficiency. The total systematic uncertainties are given by the sum in quadrature of each contributing systematic. 

\begingroup
\renewcommand*{\arraystretch}{1.2}
\begin{table}
\centering
\small
\captionof{table}{Systematic uncertainties for the \Kstarp region fit. Uncertainties are quoted as a percentage of the statistical uncertainty for a given observable, and the total uncertainty is given by the sum in quadrature of each contribution. \label{tab:systematics_Kst_Y}}
\begin{tabular}{l c c c c c c}
\hline
Observable & \textit{Eff} & \textit{PDF} & \textit{Cls} & \textit{Asym} & \textit{PID} & Total \\ \hline
$A_{\rm{SS}}^{D\pi}$ & \phantom{0}0.0 & \phantom{0}0.5 & \phantom{0}0.4 & 25.6 & \phantom{0}0.8 & 25.6 \\
$A_{\rm{OS}}^{D\pi}$ & \phantom{0}0.0 & \phantom{0}0.4 & \phantom{0}0.7 & 16.9 & \phantom{0}0.9 & 16.9 \\
$A_{\rm{SS}}^{DK}$ & \phantom{0}0.0 & \phantom{0}1.7 & 10.1 & 11.9 & \phantom{0}6.3 & 16.9 \\
$A_{\rm{OS}}^{DK}$ & \phantom{0}0.0 & \phantom{0}0.3 & 16.7 & \phantom{0}1.3 & \phantom{0}5.5 & 17.7 \\
$R_{\rm{SS/OS}}$ & 33.6 & \phantom{0}0.5 & \phantom{0}0.2 & \phantom{0}0.1 & \phantom{0}0.5 & 33.6 \\
$R_{\rm{SS}}^{DK/D\pi}$ & 29.2 & \phantom{0}3.2 & 31.3 & \phantom{0}0.1 & \phantom{0}8.1 & 43.7 \\
$R_{\rm{OS}}^{DK/D\pi}$ & 15.5 & \phantom{0}2.7 & 40.9 & \phantom{0}0.1 & \phantom{0}4.9 & 44.1 \\
\hline
\end{tabular}
\end{table}
\endgroup

\begingroup
\renewcommand*{\arraystretch}{1.2}
\begin{table}[h!]
\centering
\small
\captionof{table}{Systematic uncertainties for the non-\Kstarp region fit. Uncertainties are quoted as a percentage of the statistical uncertainty for a given observable, and the total uncertainty is given by the sum in quadrature of each contribution. \label{tab:systematics_Kst_N}}
\begin{tabular}{l c c c c c c}
\hline
Observable & \textit{Eff} & \textit{PDF} & \textit{Cls} & \textit{Asym} & \textit{PID} & Total \\ \hline
$A_{\rm{SS}}^{D\pi}$ & \phantom{0}0.0 & \phantom{0}0.4 & \phantom{0}0.6 & 14.3 & \phantom{0}1.0 & 14.4 \\
$A_{\rm{OS}}^{D\pi}$ & \phantom{0}0.0 & \phantom{0}0.7 & \phantom{0}0.5 & 18.4 & \phantom{0}1.7 & 18.5 \\
$A_{\rm{SS}}^{DK}$ & \phantom{0}0.1 & \phantom{0}0.5 & 17.8 & \phantom{0}7.1 & \phantom{0}5.8 & 20.0 \\
$A_{\rm{OS}}^{DK}$ & \phantom{0}0.0 & \phantom{0}1.5 & 10.9 & \phantom{0}1.3 & \phantom{0}9.4 & 14.5 \\
$R_{\rm{SS/OS}}$ & 48.6 & \phantom{0}0.6 & \phantom{0}0.5 & \phantom{0}0.1 & \phantom{0}0.4 & 48.6 \\
$R_{\rm{SS}}^{DK/D\pi}$ & 14.8 & \phantom{0}3.0 & 44.4 & \phantom{0}0.1 & \phantom{0}4.4 & 47.1 \\
$R_{\rm{OS}}^{DK/D\pi}$ & 18.6 & \phantom{0}4.0 & 32.7 & \phantom{0}0.1 & \phantom{0}7.3 & 38.5 \\
\hline
\end{tabular}
\end{table}
\endgroup

\FloatBarrier

\section{Results}
\label{sec:results}

The results for the \Kstarp region of the Dalitz plot are
\begin{align*}
A_{\rm{SS}}^{D\pi} &= -0.020 \pm 0.011 \pm 0.003\,, \\ 
A_{\rm{OS}}^{D\pi} &= \phantom{-}0.007 \pm 0.017 \pm 0.003\,, \\ 
A_{\rm{SS}}^{DK} &= \phantom{-}0.084 \pm 0.049 \pm 0.008\,, \\ 
A_{\rm{OS}}^{DK} &= \phantom{-}0.021 \pm 0.094 \pm 0.017\,, \\ 
R_{\rm{SS/OS}} &= \phantom{-}2.585 \pm 0.057 \pm 0.019\,, \\ 
R_{\rm{SS}}^{DK/D\pi} &= \phantom{-}0.079 \pm 0.004 \pm 0.002\,, \\ 
R_{\rm{OS}}^{DK/D\pi} &= \phantom{-}0.062 \pm 0.006 \pm 0.003 \,, 
\end{align*}
and the results for the non-\Kstarp region are
\begin{align*}
A_{\rm{SS}}^{D\pi} &= -0.034 \pm 0.020 \pm 0.003\,, \\ 
A_{\rm{OS}}^{D\pi} &= \phantom{-}0.003 \pm 0.015 \pm 0.003\,, \\ 
A_{\rm{SS}}^{DK} &= \phantom{-}0.095 \pm 0.089 \pm 0.018\,, \\ 
A_{\rm{OS}}^{DK} &= -0.038 \pm 0.075 \pm 0.011\,, \\ 
R_{\rm{SS/OS}} &= \phantom{-}0.706 \pm 0.019 \pm 0.009\,, \\ 
R_{\rm{SS}}^{DK/D\pi} &= \phantom{-}0.081 \pm 0.008 \pm 0.004\,, \\ 
R_{\rm{OS}}^{DK/D\pi} &= \phantom{-}0.073 \pm 0.006 \pm 0.002 \,.
\end{align*}

The results are in agreement with Ref.~\cite{LHCb-PAPER-2013-068}, and all statistical uncertainties are reduced in accordance with the increased signal yields. The systematic uncertainties on each asymmetry are reduced considerably due to improved knowledge of the $\Bpm$ production asymmetry and the kaon detection asymmetry. The systematic uncertainties on $R_{SS}^{DK/D\pi}$, $R_{OS}^{DK/D\pi}$, and $R_{SS/OS}$ are also reduced, due to the use of larger simulated samples. All observables are statistically limited with the current data set. The statistical and systematic correlation matrices for the \CP observables are given in App.~\ref{sec:corr_matrices}.

A comparison of the \Kstarp region results with the SM expectation is made by calculating the \CP observables from the current best-fit values of $\g=(74.0^{+5.0}_{-5.8})^\circ$, $\delta_B=(131.2^{+5.1}_{-5.9})^\circ$, and $r_B=(9.89^{+0.51}_{-0.50})\%$ for $\Bp \to D\Kp$ decays~\cite{LHCb-CONF-2018-002}; no comparison is made using the non-\Kstarp results, since the required charm hadronic parameters have not yet been measured.
For $\Bp \to D\pip$ decays, where no independent information on $r_B^\pi$ and $\delta_B^\pi$ is available, the uniform PDFs $180^\circ<\delta_B^\pi<360^\circ$ and $r_B^\pi<0.02$ are used.
The \D-decay parameters are taken from the literature: 
$r_D^2= 0.655 \pm 0.007$ and $\delta_D = (-16.6 \pm 18.4)^\circ$~\cite{LHCb-PAPER-2015-026}; $\kappa = 0.94 \pm 0.12$~\cite{CLEOKsKPi}. The small corrections due to \D mixing are not considered.

\begin{figure}[!t]
  \begin{center}
   \includegraphics[width=0.48\linewidth]{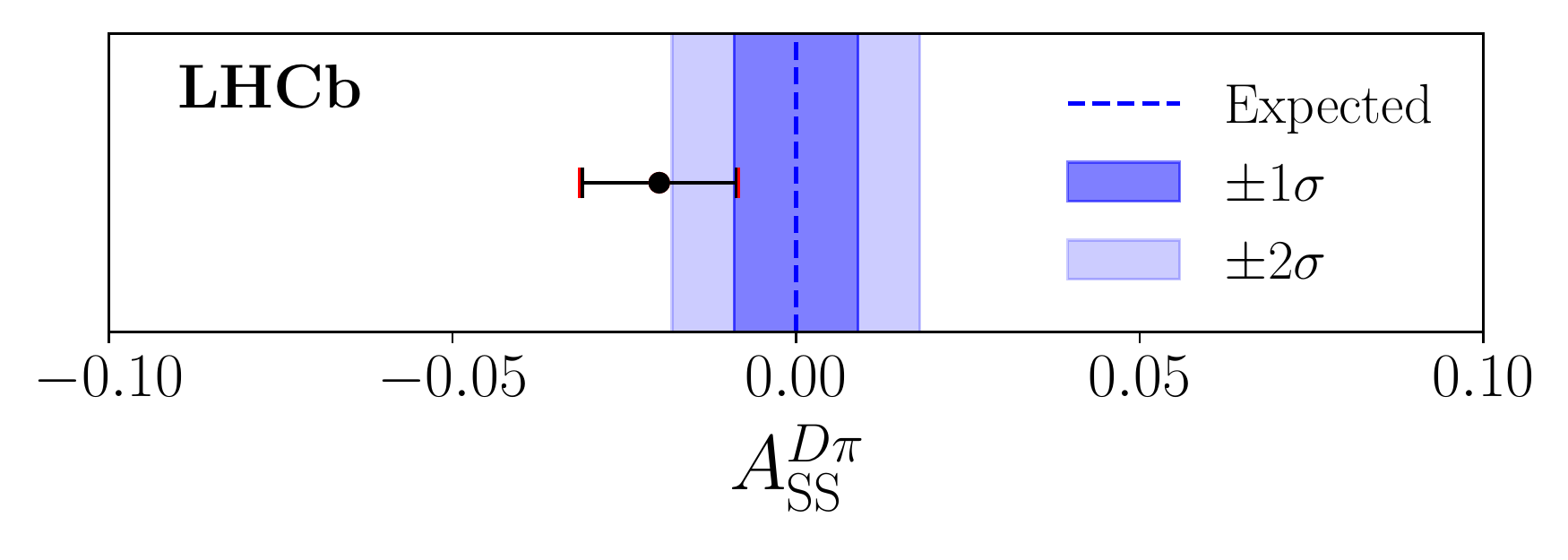}
   \includegraphics[width=0.48\linewidth]{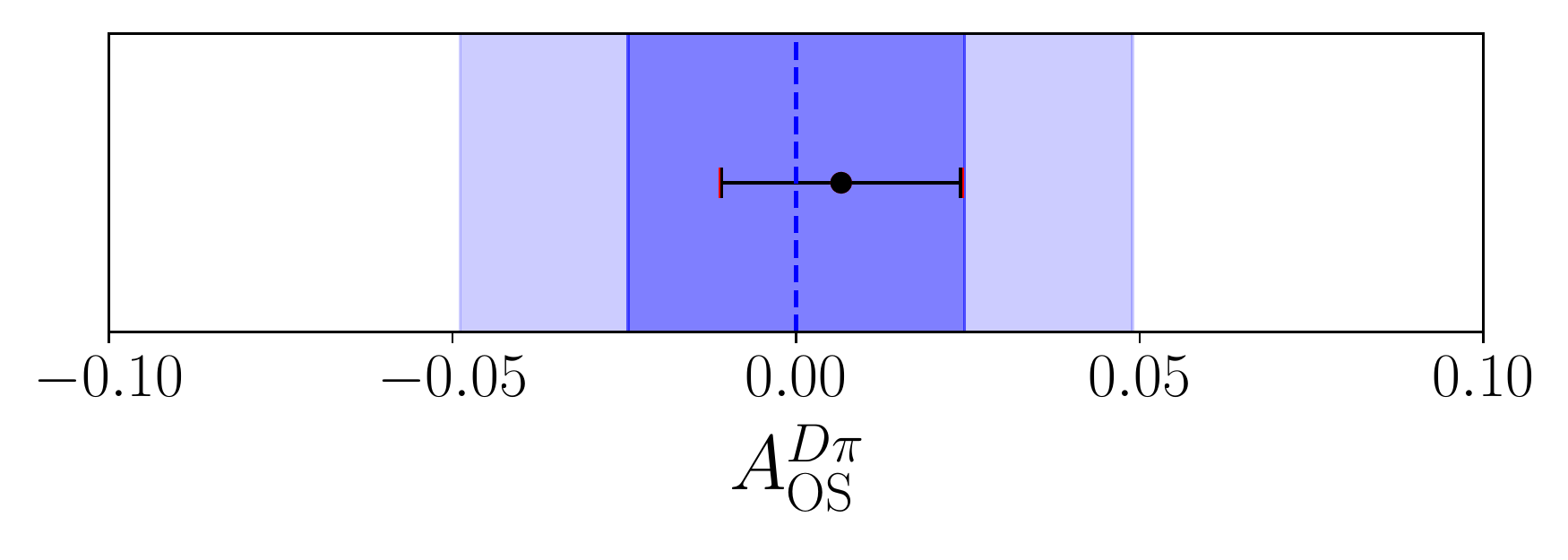}\\
   \includegraphics[width=0.48\linewidth]{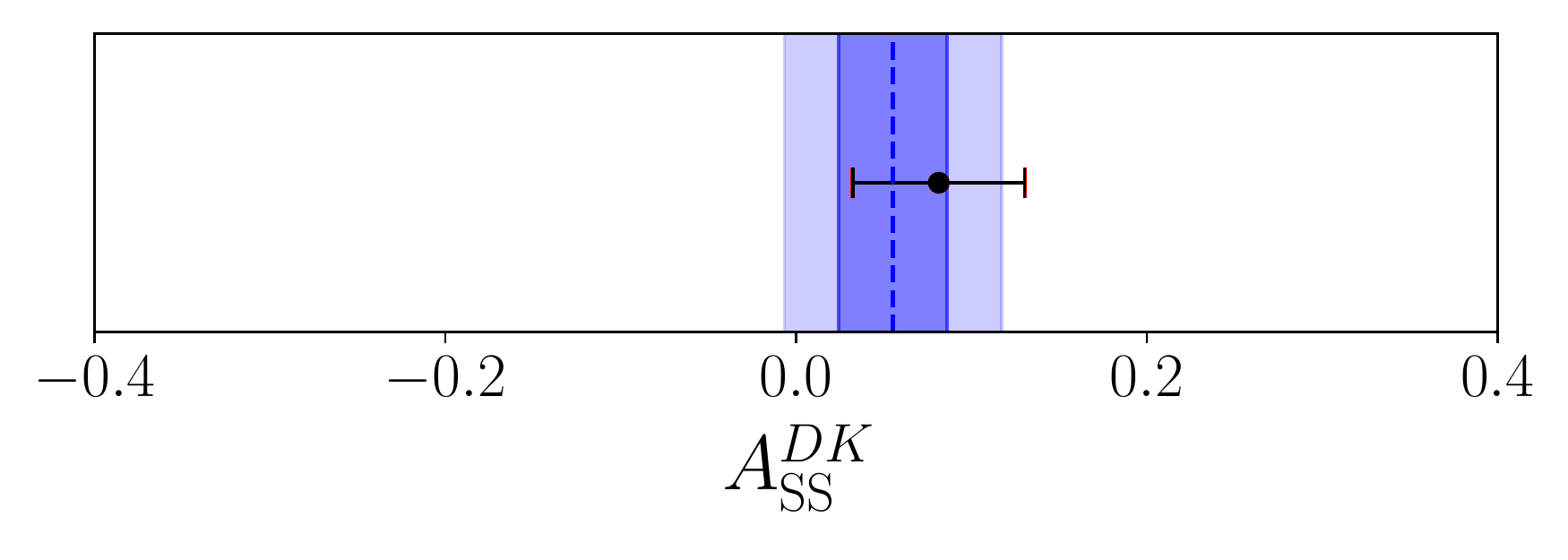}
   \includegraphics[width=0.48\linewidth]{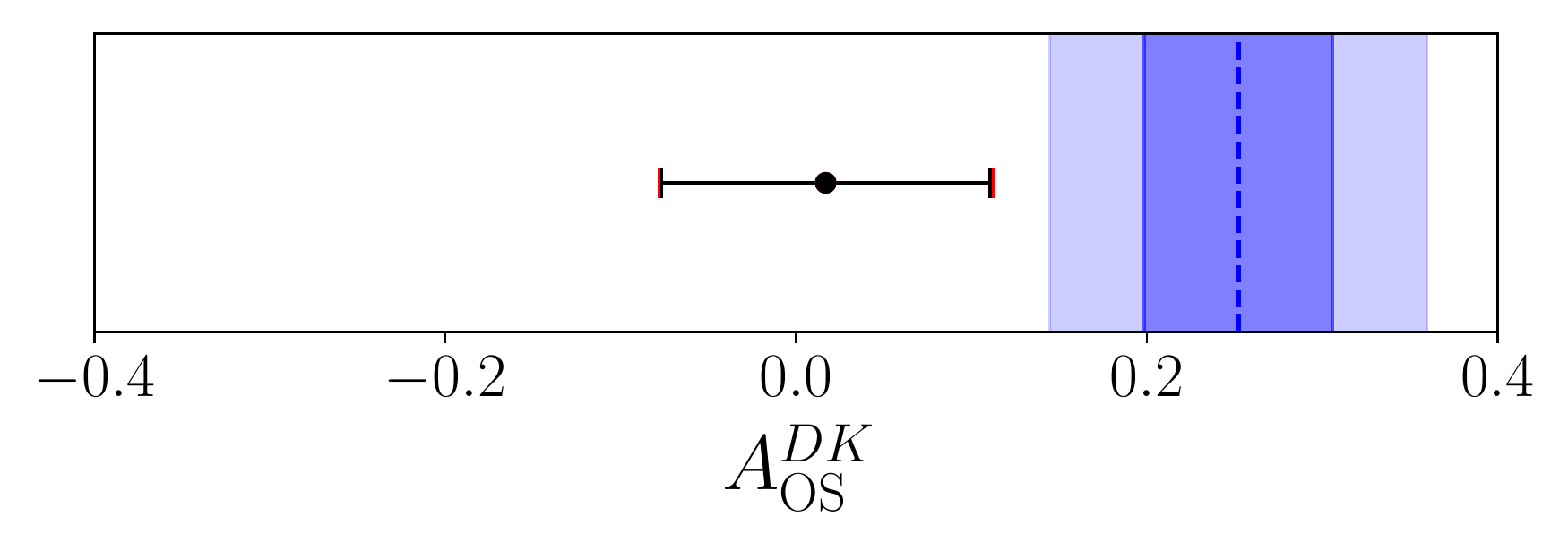}\\
   \includegraphics[width=0.48\linewidth]{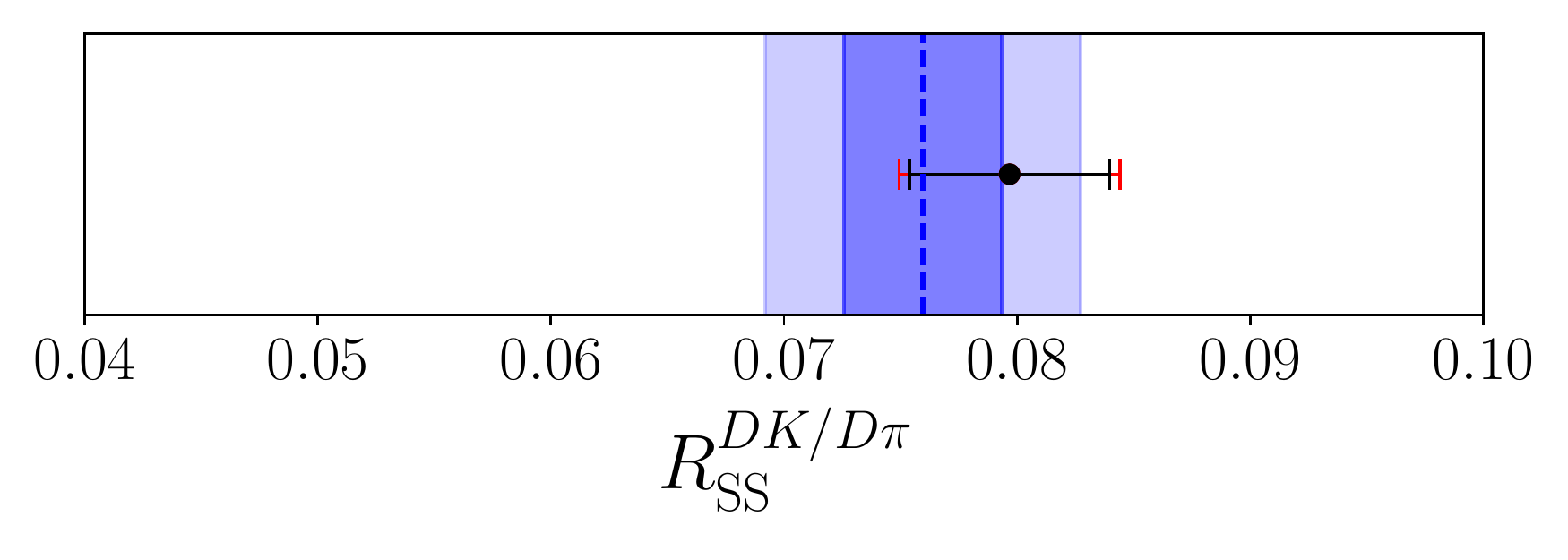}
   \includegraphics[width=0.48\linewidth]{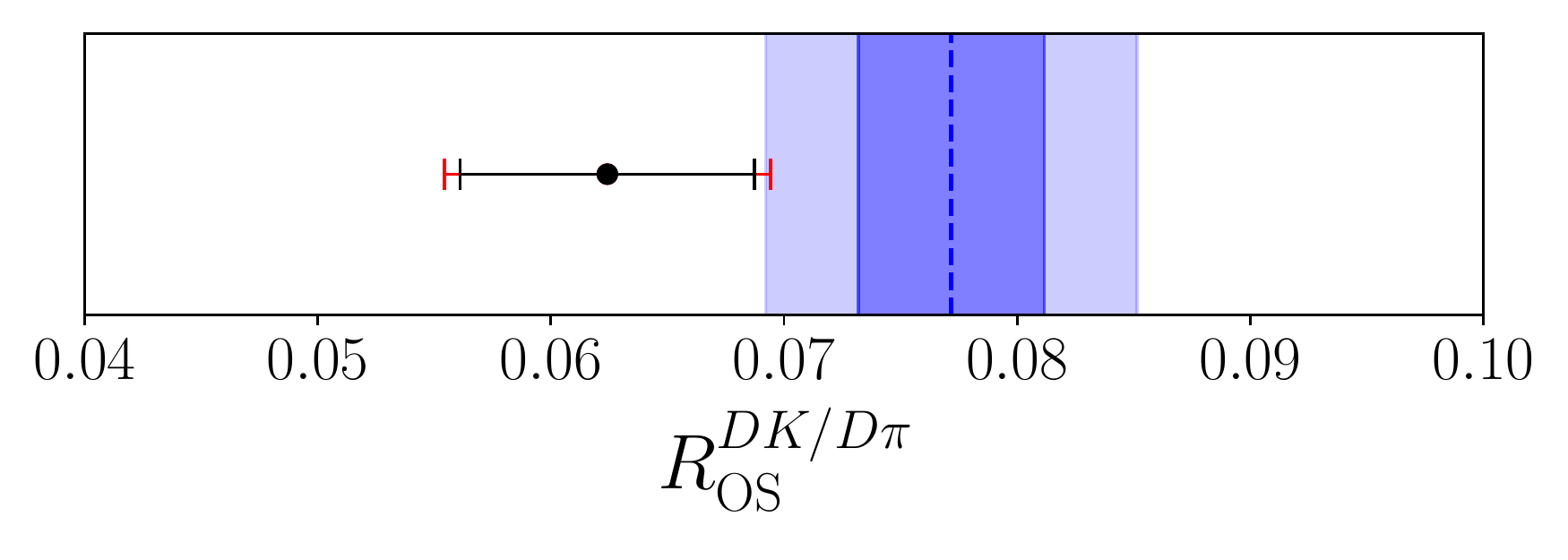}\\
   \includegraphics[width=0.48\linewidth]{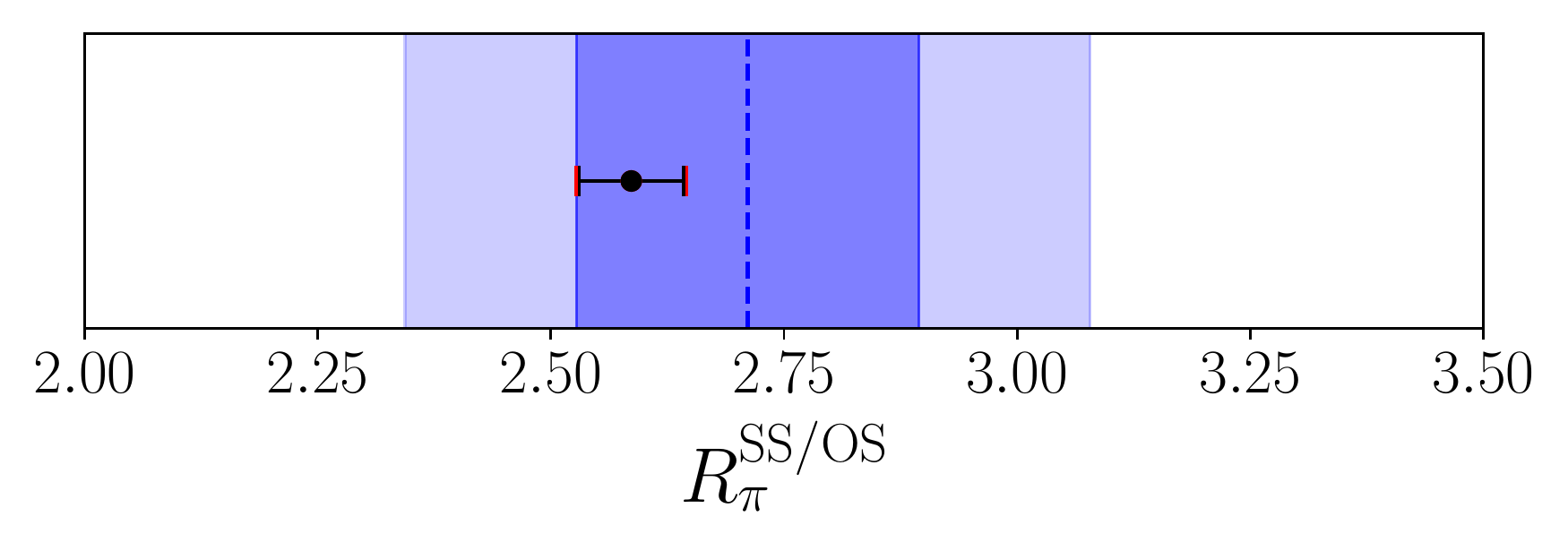}
    \end{center}
  \caption{Comparison with SM expectations for results within the \Kstarp region, using current world-average parameter values. The dashed blue line indicates the expected SM value, and the shaded dark (light) blue regions indicate the 68\% (95\%) confidence-level intervals. The results are shown as black points with black (red) error bars indicating the statistical (total) uncertainty.
  \label{fig:comparison_Kst}}
\end{figure}

For these inputs, the $68\%$ and $95\%$ confidence-level expectation intervals are displayed in Fig.~\ref{fig:comparison_Kst}, together with the results presented herein. The dominant uncertainty contribution to the expectation intervals comes from the \D-decay parameter inputs. The measurements are found to be compatible with the SM expectation, where the $\chi^2$ per degree of freedom is found to be 1.56 taking into account the uncertainties and correlations of both the measurements and the expected values; the corresponding $p$-value for rejection of the SM hypothesis is 0.14. 

\section{Conclusion}
\label{sec:conclusion}

Measurements of \CP observables in $\Bp \to D \Kp$ and $\Bp \to D \pip$ decays with the $D$ meson decaying to $\KS \Kp \pim$ and $\KS \Km \pip$ are performed using \lhcb data collected in Run 1 and Run 2. The results are in agreement with the SM, and supersede those of the previous study~\cite{LHCb-PAPER-2013-068}, benefiting from the increased data sample and improved analysis methods. The measurements presented in this paper improve the precision of several of the \CP observables used in global fits for $\gamma$, which will contribute to improved precision on $\gamma$ and on the hadronic parameters $r_B$ and $\delta_B$ for these decays. Improved measurements of charm hadronic parameters in both the \Kstarp and non-\Kstarp regions would also benefit the interpretation of these results and the constraints on $\gamma$ that can be obtained from them.

\FloatBarrier








\section*{Acknowledgements}
%
%
\noindent We express our gratitude to our colleagues in the CERN
accelerator departments for the excellent performance of the LHC. We
thank the technical and administrative staff at the LHCb
institutes.
We acknowledge support from CERN and from the national agencies:
CAPES, CNPq, FAPERJ and FINEP (Brazil); 
MOST and NSFC (China); 
CNRS/IN2P3 (France); 
BMBF, DFG and MPG (Germany); 
INFN (Italy); 
NWO (Netherlands); 
MNiSW and NCN (Poland); 
MEN/IFA (Romania); 
MSHE (Russia); 
MinECo (Spain); 
SNSF and SER (Switzerland); 
NASU (Ukraine); 
STFC (United Kingdom); 
DOE NP and NSF (USA).
We acknowledge the computing resources that are provided by CERN, IN2P3
(France), KIT and DESY (Germany), INFN (Italy), SURF (Netherlands),
PIC (Spain), GridPP (United Kingdom), RRCKI and Yandex
LLC (Russia), CSCS (Switzerland), IFIN-HH (Romania), CBPF (Brazil),
PL-GRID (Poland) and OSC (USA).
We are indebted to the communities behind the multiple open-source
software packages on which we depend.
Individual groups or members have received support from
AvH Foundation (Germany);
EPLANET, Marie Sk\l{}odowska-Curie Actions and ERC (European Union);
ANR, Labex P2IO and OCEVU, and R\'{e}gion Auvergne-Rh\^{o}ne-Alpes (France);
Key Research Program of Frontier Sciences of CAS, CAS PIFI, and the Thousand Talents Program (China);
RFBR, RSF and Yandex LLC (Russia);
GVA, XuntaGal and GENCAT (Spain);
the Royal Society
and the Leverhulme Trust (United Kingdom).

\section*{Appendices}

\appendix

\section{Correlation matrices}
\label{sec:corr_matrices}

Statistical and systematic correlation matrices for the seven \CP observables are given in Tabs.~\ref{tab:stat_corr_Y}$-$\ref{tab:syst_corr_Kst_N}, for both the $K^*(892)^\pm$ region and non-$K^*(892)^\pm$ region results.

\begin{table}[!h]

\centering

\captionof{table}{Statistical correlation matrix for the restricted $K^*(892)^\pm$ region fit. \label{tab:stat_corr_Y}}

\begin{tabular}{l |ccccccc}

& $A_{\rm{SS}}^{D\pi}$& $A_{\rm{OS}}^{D\pi}$& $A_{\rm{SS}}^{DK}$& $A_{\rm{OS}}^{DK}$& $R_{\rm{SS/OS}}$& $R_{\rm{SS}}^{DK/D\pi}$& $R_{\rm{OS}}^{DK/D\pi}$\\

\hline

$A_{\rm{SS}}^{D\pi}$ & \phantom{$-$}1 & \phantom{$-$}0.00 & $-$0.05 & \phantom{$-$}0.00 & \phantom{$-$}0.00 & $-$0.01 & \phantom{$-$}0.00\\

$A_{\rm{OS}}^{D\pi}$ & \phantom{$-$}0.00 & \phantom{$-$}1 & \phantom{$-$}0.00 & $-$0.05 & \phantom{$-$}0.00 & \phantom{$-$}0.00 & \phantom{$-$}0.00\\

$A_{\rm{SS}}^{DK}$ & $-$0.05 & \phantom{$-$}0.00 & \phantom{$-$}1 & \phantom{$-$}0.00 & \phantom{$-$}0.00 & \phantom{$-$}0.00 & \phantom{$-$}0.00\\

$A_{\rm{OS}}^{DK}$ & \phantom{$-$}0.00 & $-$0.05 & \phantom{$-$}0.00 & \phantom{$-$}1 & \phantom{$-$}0.00 & \phantom{$-$}0.00 & $-$0.02\\

$R_{\rm{SS/OS}}$ & \phantom{$-$}0.00 & \phantom{$-$}0.00 & \phantom{$-$}0.00 & \phantom{$-$}0.00 & \phantom{$-$}1 & $-$0.11 & \phantom{$-$}0.15\\

$R_{\rm{SS}}^{DK/D\pi}$ & $-$0.01 & \phantom{$-$}0.00 & \phantom{$-$}0.00 & \phantom{$-$}0.00 & $-$0.11 & \phantom{$-$}1 & \phantom{$-$}0.06\\

$R_{\rm{OS}}^{DK/D\pi}$ & \phantom{$-$}0.00 & \phantom{$-$}0.00 & \phantom{$-$}0.00 & $-$0.02 & \phantom{$-$}0.15 & \phantom{$-$}0.06 & \phantom{$-$}1\\

\hline

\end{tabular}

\end{table}

\begin{table}[!h]
\centering
\captionof{table}{Systematic correlation matrix for the $K^*(892)^\pm$ region fit. \label{tab:syst_corr_Kst_Y}}
\begin{tabular}{l | c c c c c c c}
& $A_{\rm{SS}}^{D\pi}$ & $A_{\rm{OS}}^{D\pi}$ & $A_{\rm{SS}}^{DK}$ & $A_{\rm{OS}}^{DK}$ & $R_{\rm{SS/OS}}$ & $R_{\rm{SS}}^{DK/D\pi}$ & $R_{\rm{OS}}^{DK/D\pi}$ \\
\hline
$A_{\rm{SS}}^{D\pi}$ & \phantom{$-$}1 & $-$0.88 & \phantom{$-$}0.74 & \phantom{$-$}0.05 & \phantom{$-$}0.00 & \phantom{$-$}0.00 & \phantom{$-$}0.00\\
$A_{\rm{OS}}^{D\pi}$ & $-$0.88 & \phantom{$-$}1 & $-$0.73 & $-$0.08 & \phantom{$-$}0.00 & \phantom{$-$}0.00 & \phantom{$-$}0.00\\
$A_{\rm{SS}}^{DK}$ & \phantom{$-$}0.74 & $-$0.73 & \phantom{$-$}1 & \phantom{$-$}0.63 & \phantom{$-$}0.00 & $-$0.17 & \phantom{$-$}0.00\\
$A_{\rm{OS}}^{DK}$ & \phantom{$-$}0.05 & $-$0.08 & \phantom{$-$}0.63 & \phantom{$-$}1 & \phantom{$-$}0.00 & \phantom{$-$}0.00 & $-$0.10\\
$R_{\rm{SS/OS}}$ & \phantom{$-$}0.00 & \phantom{$-$}0.00 & \phantom{$-$}0.00 & \phantom{$-$}0.00 & \phantom{$-$}1 & \phantom{$-$}0.00 & \phantom{$-$}0.00\\
$R_{\rm{SS}}^{DK/D\pi}$ & \phantom{$-$}0.00 & \phantom{$-$}0.00 & $-$0.17 & \phantom{$-$}0.00 & \phantom{$-$}0.00 & \phantom{$-$}1 & \phantom{$-$}0.25\\
$R_{\rm{OS}}^{DK/D\pi}$ & \phantom{$-$}0.00 & \phantom{$-$}0.00 & \phantom{$-$}0.00 & $-$0.10 & \phantom{$-$}0.00 & \phantom{$-$}0.25 & \phantom{$-$}1\\
\hline
\end{tabular}
\end{table}

\begin{table}[!h]

\centering

\captionof{table}{Statistical correlation matrix for the non$-$$K^*(892)^\pm$ region fit. \label{tab:stat_corr_N}}

\begin{tabular}{l |ccccccc}

& $A_{\rm{SS}}^{D\pi}$& $A_{\rm{OS}}^{D\pi}$& $A_{\rm{SS}}^{DK}$& $A_{\rm{OS}}^{DK}$& $R_{\rm{SS/OS}}$& $R_{\rm{SS}}^{DK/D\pi}$& $R_{\rm{OS}}^{DK/D\pi}$\\

\hline

$A_{\rm{SS}}^{D\pi}$ & \phantom{$-$}1 & \phantom{$-$}0.00 & $-$0.03 & \phantom{$-$}0.00 & \phantom{$-$}0.01 & $-$0.01 & \phantom{$-$}0.00\\

$A_{\rm{OS}}^{D\pi}$ & \phantom{$-$}0.00 & \phantom{$-$}1 & \phantom{$-$}0.00 & $-$0.05 & \phantom{$-$}0.00 & \phantom{$-$}0.00 & \phantom{$-$}0.00\\

$A_{\rm{SS}}^{DK}$ & $-$0.03 & \phantom{$-$}0.00 & \phantom{$-$}1 & \phantom{$-$}0.00 & \phantom{$-$}0.00 & $-$0.03 & \phantom{$-$}0.00\\

$A_{\rm{OS}}^{DK}$ & \phantom{$-$}0.00 & $-$0.05 & \phantom{$-$}0.00 & \phantom{$-$}1 & \phantom{$-$}0.00 & $-$0.01 & $-$0.01\\

$R_{\rm{SS/OS}}$ & \phantom{$-$}0.01 & \phantom{$-$}0.00 & \phantom{$-$}0.00 & \phantom{$-$}0.00 & \phantom{$-$}1 & $-$0.16 & \phantom{$-$}0.11\\

$R_{\rm{SS}}^{DK/D\pi}$ & $-$0.01 & \phantom{$-$}0.00 & $-$0.03 & $-$0.01 & $-$0.16 & \phantom{$-$}1 & \phantom{$-$}0.09\\

$R_{\rm{OS}}^{DK/D\pi}$ & \phantom{$-$}0.00 & \phantom{$-$}0.00 & \phantom{$-$}0.00 & $-$0.01 & \phantom{$-$}0.11 & \phantom{$-$}0.09 & \phantom{$-$}1\\

\hline

\end{tabular}

\end{table}

\begin{table}[!h]
\centering
\captionof{table}{Systematic correlation matrix for the non$-$$K^*(892)^\pm$ region fit. \label{tab:syst_corr_Kst_N}}
\begin{tabular}{l | c c c c c c c}
& $A_{\rm{SS}}^{D\pi}$ & $A_{\rm{OS}}^{D\pi}$ & $A_{\rm{SS}}^{DK}$ & $A_{\rm{OS}}^{DK}$ & $R_{\rm{SS/OS}}$ & $R_{\rm{SS}}^{DK/D\pi}$ & $R_{\rm{OS}}^{DK/D\pi}$ \\
\hline
$A_{\rm{SS}}^{D\pi}$ & \phantom{$-$}1 & $-$0.84 & \phantom{$-$}0.31 & \phantom{$-$}0.05 & \phantom{$-$}0.00 & \phantom{$-$}0.00 & \phantom{$-$}0.00\\
$A_{\rm{OS}}^{D\pi}$ & $-$0.84 & \phantom{$-$}1 & $-$0.35 & $-$0.06 & \phantom{$-$}0.00 & \phantom{$-$}0.00 & \phantom{$-$}0.00\\
$A_{\rm{SS}}^{DK}$ & \phantom{$-$}0.31 & $-$0.35 & \phantom{$-$}1 & \phantom{$-$}0.81 & \phantom{$-$}0.00 & $-$0.28 & \phantom{$-$}0.00\\
$A_{\rm{OS}}^{DK}$ & \phantom{$-$}0.05 & $-$0.06 & \phantom{$-$}0.81 & \phantom{$-$}1 & \phantom{$-$}0.00 & \phantom{$-$}0.00 & \phantom{$-$}0.03\\
$R_{\rm{SS/OS}}$ & \phantom{$-$}0.00 & \phantom{$-$}0.00 & \phantom{$-$}0.00 & \phantom{$-$}0.00 & \phantom{$-$}1 & \phantom{$-$}0.00 & \phantom{$-$}0.00\\
$R_{\rm{SS}}^{DK/D\pi}$ & \phantom{$-$}0.00 & \phantom{$-$}0.00 & $-$0.28 & \phantom{$-$}0.00 & \phantom{$-$}0.00 & \phantom{$-$}1 & \phantom{$-$}0.02\\
$R_{\rm{OS}}^{DK/D\pi}$ & \phantom{$-$}0.00 & \phantom{$-$}0.00 & \phantom{$-$}0.00 & \phantom{$-$}0.03 & \phantom{$-$}0.00 & \phantom{$-$}0.02 & \phantom{$-$}1\\
\hline
\end{tabular}
\end{table}

\FloatBarrier


\addcontentsline{toc}{section}{References}
\bibliographystyle{LHCb}
\bibliography{main,standard,LHCb-PAPER,LHCb-CONF,LHCb-DP,LHCb-TDR}

\ifx\mcitethebibliography\mciteundefinedmacro
\PackageError{LHCb.bst}{mciteplus.sty has not been loaded}
{This bibstyle requires the use of the mciteplus package.}\fi
\providecommand{\href}[2]{#2}
\begin{mcitethebibliography}{10}
\mciteSetBstSublistMode{n}
\mciteSetBstMaxWidthForm{subitem}{\alph{mcitesubitemcount})}
\mciteSetBstSublistLabelBeginEnd{\mcitemaxwidthsubitemform\space}
{\relax}{\relax}

\bibitem{Cabibbo:1963yz}
N.~Cabibbo, \ifthenelse{\boolean{articletitles}}{\emph{{Unitary symmetry and
  leptonic decays}},
  }{}\href{https://doi.org/10.1103/PhysRevLett.10.531}{Phys.\ Rev.\ Lett.\
  \textbf{10} (1963) 531}\relax
\mciteBstWouldAddEndPuncttrue
\mciteSetBstMidEndSepPunct{\mcitedefaultmidpunct}
{\mcitedefaultendpunct}{\mcitedefaultseppunct}\relax
\EndOfBibitem
\bibitem{Kobayashi:1973fv}
M.~Kobayashi and T.~Maskawa,
  \ifthenelse{\boolean{articletitles}}{\emph{{\CP-violation in the
  renormalizable theory of weak interaction}},
  }{}\href{https://doi.org/10.1143/PTP.49.652}{Prog.\ Theor.\ Phys.\
  \textbf{49} (1973) 652}\relax
\mciteBstWouldAddEndPuncttrue
\mciteSetBstMidEndSepPunct{\mcitedefaultmidpunct}
{\mcitedefaultendpunct}{\mcitedefaultseppunct}\relax
\EndOfBibitem
\bibitem{wolfenstein}
L.~Wolfenstein, \ifthenelse{\boolean{articletitles}}{\emph{Parametrization of
  the {K}obayashi--{M}askawa matrix},
  }{}\href{https://doi.org/10.1103/PhysRevLett.51.1945}{Phys.\ Rev.\ Lett.\
  \textbf{51} (1983) 1945}\relax
\mciteBstWouldAddEndPuncttrue
\mciteSetBstMidEndSepPunct{\mcitedefaultmidpunct}
{\mcitedefaultendpunct}{\mcitedefaultseppunct}\relax
\EndOfBibitem
\bibitem{King:2019rvk}
D.~King, M.~Kirk, A.~Lenz, and T.~Rauh,
  \ifthenelse{\boolean{articletitles}}{\emph{{$|V_{cb}|$ and $\gamma$ from
  $B$-mixing}}, }{}\href{http://arxiv.org/abs/1911.07856}{{\normalfont\ttfamily
  arXiv:1911.07856}}\relax
\mciteBstWouldAddEndPuncttrue
\mciteSetBstMidEndSepPunct{\mcitedefaultmidpunct}
{\mcitedefaultendpunct}{\mcitedefaultseppunct}\relax
\EndOfBibitem
\bibitem{LHCb-CONF-2018-002}
{LHCb collaboration}, \ifthenelse{\boolean{articletitles}}{\emph{{Update of the
  LHCb combination of the CKM angle $\gamma$ using $B \to DK$ decays}}, }{}
  \href{http://cdsweb.cern.ch/search?p=LHCb-CONF-2018-002&f=reportnumber&action_search=Search&c=LHCb+Conference+Contributions}
  {LHCb-CONF-2018-002}, {2018}\relax
\mciteBstWouldAddEndPuncttrue
\mciteSetBstMidEndSepPunct{\mcitedefaultmidpunct}
{\mcitedefaultendpunct}{\mcitedefaultseppunct}\relax
\EndOfBibitem
\bibitem{LHCb-PAPER-2016-032}
LHCb collaboration, R.~Aaij {\em et~al.},
  \ifthenelse{\boolean{articletitles}}{\emph{{Measurement of the CKM angle
  $\gamma$ from a combination of LHCb results}},
  }{}\href{https://doi.org/10.1007/JHEP12(2016)087}{JHEP \textbf{12} (2016)
  087}, \href{http://arxiv.org/abs/1611.03076}{{\normalfont\ttfamily
  arXiv:1611.03076}}\relax
\mciteBstWouldAddEndPuncttrue
\mciteSetBstMidEndSepPunct{\mcitedefaultmidpunct}
{\mcitedefaultendpunct}{\mcitedefaultseppunct}\relax
\EndOfBibitem
\bibitem{LHCb-PAPER-2013-068}
LHCb collaboration, R.~Aaij {\em et~al.},
  \ifthenelse{\boolean{articletitles}}{\emph{{A study of \CP violation in
  \mbox{\decay{\Bpm}{\D\Kpm}} and \mbox{\decay{\Bpm}{\D\pipm}} decays with
  \mbox{\decay{\D}{\KS\Kpm\pimp}} final states}},
  }{}\href{https://doi.org/10.1016/j.physletb.2014.03.051}{Phys.\ Lett.\
  \textbf{B733} (2014) 36},
  \href{http://arxiv.org/abs/1402.2982}{{\normalfont\ttfamily
  arXiv:1402.2982}}\relax
\mciteBstWouldAddEndPuncttrue
\mciteSetBstMidEndSepPunct{\mcitedefaultmidpunct}
{\mcitedefaultendpunct}{\mcitedefaultseppunct}\relax
\EndOfBibitem
\bibitem{CLEOKsKPi}
CLEO collaboration, J.~Insler {\em et~al.},
  \ifthenelse{\boolean{articletitles}}{\emph{{Studies of the decays $D^0
  \rightarrow K_S^0K^-\pi^+$ and $D^0 \rightarrow K_S^0K^+\pi^-$}},
  }{}\href{https://doi.org/10.1103/PhysRevD.85.092016}{Phys.\ Rev.\
  \textbf{D85} (2012) 092016}, Erratum
  \href{https://doi.org/10.1103/PhysRevD.94.099905}{ibid.\   \textbf{D94}
  (2016) 099905}, \href{http://arxiv.org/abs/1203.3804}{{\normalfont\ttfamily
  arXiv:1203.3804}}\relax
\mciteBstWouldAddEndPuncttrue
\mciteSetBstMidEndSepPunct{\mcitedefaultmidpunct}
{\mcitedefaultendpunct}{\mcitedefaultseppunct}\relax
\EndOfBibitem
\bibitem{Grossman:2002aq}
Y.~Grossman, Z.~Ligeti, and A.~Soffer,
  \ifthenelse{\boolean{articletitles}}{\emph{{Measuring gamma in ${B^{\pm}}
  \rightarrow K^\pm (K K^{*})_{D}$ decays}},
  }{}\href{https://doi.org/10.1103/PhysRevD.67.071301}{Phys.\ Rev.\
  \textbf{D67} (2003) 071301},
  \href{http://arxiv.org/abs/hep-ph/0210433}{{\normalfont\ttfamily
  arXiv:hep-ph/0210433}}\relax
\mciteBstWouldAddEndPuncttrue
\mciteSetBstMidEndSepPunct{\mcitedefaultmidpunct}
{\mcitedefaultendpunct}{\mcitedefaultseppunct}\relax
\EndOfBibitem
\bibitem{Atwood:2003mj}
D.~Atwood and A.~Soni, \ifthenelse{\boolean{articletitles}}{\emph{{Role of
  charm factory in extracting CKM phase information via $B \rightarrow DK$}},
  }{}\href{https://doi.org/10.1103/PhysRevD.68.033003}{Phys.\ Rev.\
  \textbf{D68} (2003) 033003},
  \href{http://arxiv.org/abs/hep-ph/0304085}{{\normalfont\ttfamily
  arXiv:hep-ph/0304085}}\relax
\mciteBstWouldAddEndPuncttrue
\mciteSetBstMidEndSepPunct{\mcitedefaultmidpunct}
{\mcitedefaultendpunct}{\mcitedefaultseppunct}\relax
\EndOfBibitem
\bibitem{Atwood:1996ci}
D.~Atwood, I.~Dunietz, and A.~Soni,
  \ifthenelse{\boolean{articletitles}}{\emph{{Enhanced CP violation with $B
  \rightarrow K \Dz (\Dzb)$ modes and extraction of the CKM angle gamma}},
  }{}\href{https://doi.org/10.1103/PhysRevLett.78.3257}{Phys.\ Rev.\ Lett.\
  \textbf{78} (1997) 3257},
  \href{http://arxiv.org/abs/hep-ph/9612433}{{\normalfont\ttfamily
  arXiv:hep-ph/9612433}}\relax
\mciteBstWouldAddEndPuncttrue
\mciteSetBstMidEndSepPunct{\mcitedefaultmidpunct}
{\mcitedefaultendpunct}{\mcitedefaultseppunct}\relax
\EndOfBibitem
\bibitem{LHCb-DP-2008-001}
LHCb collaboration, A.~A. Alves~Jr.\ {\em et~al.},
  \ifthenelse{\boolean{articletitles}}{\emph{{The \lhcb detector at the LHC}},
  }{}\href{https://doi.org/10.1088/1748-0221/3/08/S08005}{JINST \textbf{3}
  (2008) S08005}\relax
\mciteBstWouldAddEndPuncttrue
\mciteSetBstMidEndSepPunct{\mcitedefaultmidpunct}
{\mcitedefaultendpunct}{\mcitedefaultseppunct}\relax
\EndOfBibitem
\bibitem{LHCb-DP-2014-002}
LHCb collaboration, R.~Aaij {\em et~al.},
  \ifthenelse{\boolean{articletitles}}{\emph{{LHCb detector performance}},
  }{}\href{https://doi.org/10.1142/S0217751X15300227}{Int.\ J.\ Mod.\ Phys.\
  \textbf{A30} (2015) 1530022},
  \href{http://arxiv.org/abs/1412.6352}{{\normalfont\ttfamily
  arXiv:1412.6352}}\relax
\mciteBstWouldAddEndPuncttrue
\mciteSetBstMidEndSepPunct{\mcitedefaultmidpunct}
{\mcitedefaultendpunct}{\mcitedefaultseppunct}\relax
\EndOfBibitem
\bibitem{BBDT}
V.~V. Gligorov and M.~Williams,
  \ifthenelse{\boolean{articletitles}}{\emph{{Efficient, reliable and fast
  high-level triggering using a bonsai boosted decision tree}},
  }{}\href{https://doi.org/10.1088/1748-0221/8/02/P02013}{JINST \textbf{8}
  (2013) P02013}, \href{http://arxiv.org/abs/1210.6861}{{\normalfont\ttfamily
  arXiv:1210.6861}}\relax
\mciteBstWouldAddEndPuncttrue
\mciteSetBstMidEndSepPunct{\mcitedefaultmidpunct}
{\mcitedefaultendpunct}{\mcitedefaultseppunct}\relax
\EndOfBibitem
\bibitem{Sjostrand:2007gs}
T.~Sj\"{o}strand, S.~Mrenna, and P.~Skands,
  \ifthenelse{\boolean{articletitles}}{\emph{{A brief introduction to PYTHIA
  8.1}}, }{}\href{https://doi.org/10.1016/j.cpc.2008.01.036}{Comput.\ Phys.\
  Commun.\  \textbf{178} (2008) 852},
  \href{http://arxiv.org/abs/0710.3820}{{\normalfont\ttfamily
  arXiv:0710.3820}}\relax
\mciteBstWouldAddEndPuncttrue
\mciteSetBstMidEndSepPunct{\mcitedefaultmidpunct}
{\mcitedefaultendpunct}{\mcitedefaultseppunct}\relax
\EndOfBibitem
\bibitem{LHCb-PROC-2010-056}
I.~Belyaev {\em et~al.}, \ifthenelse{\boolean{articletitles}}{\emph{{Handling
  of the generation of primary events in Gauss, the LHCb simulation
  framework}}, }{}\href{https://doi.org/10.1088/1742-6596/331/3/032047}{J.\
  Phys.\ Conf.\ Ser.\  \textbf{331} (2011) 032047}\relax
\mciteBstWouldAddEndPuncttrue
\mciteSetBstMidEndSepPunct{\mcitedefaultmidpunct}
{\mcitedefaultendpunct}{\mcitedefaultseppunct}\relax
\EndOfBibitem
\bibitem{Lange:2001uf}
D.~J. Lange, \ifthenelse{\boolean{articletitles}}{\emph{{The EvtGen particle
  decay simulation package}},
  }{}\href{https://doi.org/10.1016/S0168-9002(01)00089-4}{Nucl.\ Instrum.\
  Meth.\  \textbf{A462} (2001) 152}\relax
\mciteBstWouldAddEndPuncttrue
\mciteSetBstMidEndSepPunct{\mcitedefaultmidpunct}
{\mcitedefaultendpunct}{\mcitedefaultseppunct}\relax
\EndOfBibitem
\bibitem{Golonka:2005pn}
P.~Golonka and Z.~Was, \ifthenelse{\boolean{articletitles}}{\emph{{PHOTOS Monte
  Carlo: A precision tool for QED corrections in $Z$ and $W$ decays}},
  }{}\href{https://doi.org/10.1140/epjc/s2005-02396-4}{Eur.\ Phys.\ J.\
  \textbf{C45} (2006) 97},
  \href{http://arxiv.org/abs/hep-ph/0506026}{{\normalfont\ttfamily
  arXiv:hep-ph/0506026}}\relax
\mciteBstWouldAddEndPuncttrue
\mciteSetBstMidEndSepPunct{\mcitedefaultmidpunct}
{\mcitedefaultendpunct}{\mcitedefaultseppunct}\relax
\EndOfBibitem
\bibitem{Allison:2006ve}
Geant4 collaboration, J.~Allison {\em et~al.},
  \ifthenelse{\boolean{articletitles}}{\emph{{Geant4 developments and
  applications}}, }{}\href{https://doi.org/10.1109/TNS.2006.869826}{IEEE
  Trans.\ Nucl.\ Sci.\  \textbf{53} (2006) 270}\relax
\mciteBstWouldAddEndPuncttrue
\mciteSetBstMidEndSepPunct{\mcitedefaultmidpunct}
{\mcitedefaultendpunct}{\mcitedefaultseppunct}\relax
\EndOfBibitem
\bibitem{Agostinelli:2002hh}
Geant4 collaboration, S.~Agostinelli {\em et~al.},
  \ifthenelse{\boolean{articletitles}}{\emph{{Geant4: A simulation toolkit}},
  }{}\href{https://doi.org/10.1016/S0168-9002(03)01368-8}{Nucl.\ Instrum.\
  Meth.\  \textbf{A506} (2003) 250}\relax
\mciteBstWouldAddEndPuncttrue
\mciteSetBstMidEndSepPunct{\mcitedefaultmidpunct}
{\mcitedefaultendpunct}{\mcitedefaultseppunct}\relax
\EndOfBibitem
\bibitem{LHCb-PROC-2011-006}
M.~Clemencic {\em et~al.}, \ifthenelse{\boolean{articletitles}}{\emph{{The
  \lhcb simulation application, Gauss: Design, evolution and experience}},
  }{}\href{https://doi.org/10.1088/1742-6596/331/3/032023}{J.\ Phys.\ Conf.\
  Ser.\  \textbf{331} (2011) 032023}\relax
\mciteBstWouldAddEndPuncttrue
\mciteSetBstMidEndSepPunct{\mcitedefaultmidpunct}
{\mcitedefaultendpunct}{\mcitedefaultseppunct}\relax
\EndOfBibitem
\bibitem{PDG2018}
Particle Data Group, M.~Tanabashi {\em et~al.},
  \ifthenelse{\boolean{articletitles}}{\emph{{\href{http://pdg.lbl.gov/}{Review
  of particle physics}}},
  }{}\href{https://doi.org/10.1103/PhysRevD.98.030001}{Phys.\ Rev.\
  \textbf{D98} (2018) 030001}\relax
\mciteBstWouldAddEndPuncttrue
\mciteSetBstMidEndSepPunct{\mcitedefaultmidpunct}
{\mcitedefaultendpunct}{\mcitedefaultseppunct}\relax
\EndOfBibitem
\bibitem{Breiman}
L.~Breiman, J.~H. Friedman, R.~A. Olshen, and C.~J. Stone, {\em Classification
  and regression trees}, Wadsworth international group, Belmont, California,
  USA, 1984\relax
\mciteBstWouldAddEndPuncttrue
\mciteSetBstMidEndSepPunct{\mcitedefaultmidpunct}
{\mcitedefaultendpunct}{\mcitedefaultseppunct}\relax
\EndOfBibitem
\bibitem{Hulsbergen:2005pu}
W.~D. Hulsbergen, \ifthenelse{\boolean{articletitles}}{\emph{{Decay chain
  fitting with a Kalman filter}},
  }{}\href{https://doi.org/10.1016/j.nima.2005.06.078}{Nucl.\ Instrum.\ Meth.\
  \textbf{A552} (2005) 566},
  \href{http://arxiv.org/abs/physics/0503191}{{\normalfont\ttfamily
  arXiv:physics/0503191}}\relax
\mciteBstWouldAddEndPuncttrue
\mciteSetBstMidEndSepPunct{\mcitedefaultmidpunct}
{\mcitedefaultendpunct}{\mcitedefaultseppunct}\relax
\EndOfBibitem
\bibitem{CrystalBall}
T.~Skwarnicki, {\em {A study of the radiative CASCADE transitions between the
  Upsilon-Prime and Upsilon resonances}}, PhD thesis, Cracow, INP, 1986\relax
\mciteBstWouldAddEndPuncttrue
\mciteSetBstMidEndSepPunct{\mcitedefaultmidpunct}
{\mcitedefaultendpunct}{\mcitedefaultseppunct}\relax
\EndOfBibitem
\bibitem{LHCb-PAPER-2017-021}
LHCb collaboration, R.~Aaij {\em et~al.},
  \ifthenelse{\boolean{articletitles}}{\emph{{Measurement of \CP observables in
  \mbox{\decay{\Bpm}{D^{(\ast)}\Kpm}} and \mbox{\decay{\Bpm}{D^{(\ast)}\pipm}}
  decays}}, }{}\href{https://doi.org/10.1016/j.physletb.2017.11.070}{Phys.\
  Lett.\  \textbf{B777} (2018) 16},
  \href{http://arxiv.org/abs/1708.06370}{{\normalfont\ttfamily
  arXiv:1708.06370}}\relax
\mciteBstWouldAddEndPuncttrue
\mciteSetBstMidEndSepPunct{\mcitedefaultmidpunct}
{\mcitedefaultendpunct}{\mcitedefaultseppunct}\relax
\EndOfBibitem
\bibitem{LHCb-PAPER-2014-036}
LHCb collaboration, R.~Aaij {\em et~al.},
  \ifthenelse{\boolean{articletitles}}{\emph{{Dalitz plot analysis of
  \mbox{\decay{\Bs}{\Dzb\Km\pip}} decays}},
  }{}\href{https://doi.org/10.1103/PhysRevD.90.072003}{Phys.\ Rev.\
  \textbf{D90} (2014) 072003},
  \href{http://arxiv.org/abs/1407.7712}{{\normalfont\ttfamily
  arXiv:1407.7712}}\relax
\mciteBstWouldAddEndPuncttrue
\mciteSetBstMidEndSepPunct{\mcitedefaultmidpunct}
{\mcitedefaultendpunct}{\mcitedefaultseppunct}\relax
\EndOfBibitem
\bibitem{LHCb-DP-2018-001}
R.~Aaij {\em et~al.}, \ifthenelse{\boolean{articletitles}}{\emph{{Selection and
  processing of calibration samples to measure the particle identification
  performance of the LHCb experiment in Run 2}},
  }{}\href{https://doi.org/10.1140/epjti/s40485-019-0050-z}{Eur.\ Phys.\ J.\
  Tech.\ Instr.\  \textbf{6} (2018) 1},
  \href{http://arxiv.org/abs/1803.00824}{{\normalfont\ttfamily
  arXiv:1803.00824}}\relax
\mciteBstWouldAddEndPuncttrue
\mciteSetBstMidEndSepPunct{\mcitedefaultmidpunct}
{\mcitedefaultendpunct}{\mcitedefaultseppunct}\relax
\EndOfBibitem
\bibitem{LHCb-PUB-2018-004}
A.~Davis {\em et~al.}, \ifthenelse{\boolean{articletitles}}{\emph{{Measurement
  of the instrumental asymmetry for $K^{-}\pi^{+}$-pairs at LHCb in Run 2}},
  }{}
  \href{http://cdsweb.cern.ch/search?p=LHCb-PUB-2018-004&f=reportnumber&action_search=Search&c=LHCb+Notes}
  {LHCb-PUB-2018-004}, 2018\relax
\mciteBstWouldAddEndPuncttrue
\mciteSetBstMidEndSepPunct{\mcitedefaultmidpunct}
{\mcitedefaultendpunct}{\mcitedefaultseppunct}\relax
\EndOfBibitem
\bibitem{LHCb-PAPER-2016-054}
LHCb collaboration, R.~Aaij {\em et~al.},
  \ifthenelse{\boolean{articletitles}}{\emph{{Measurement of the \Bpm
  production asymmetry and the \CP asymmetry in \mbox{\decay{\Bpm}{\jpsi \Kpm}}
  decays}}, }{}\href{https://doi.org/10.1103/PhysRevD.95.052005}{Phys.\ Rev.\
  \textbf{D95} (2017) 052005},
  \href{http://arxiv.org/abs/1701.05501}{{\normalfont\ttfamily
  arXiv:1701.05501}}\relax
\mciteBstWouldAddEndPuncttrue
\mciteSetBstMidEndSepPunct{\mcitedefaultmidpunct}
{\mcitedefaultendpunct}{\mcitedefaultseppunct}\relax
\EndOfBibitem
\bibitem{LHCb-PAPER-2015-026}
LHCb collaboration, R.~Aaij {\em et~al.},
  \ifthenelse{\boolean{articletitles}}{\emph{{Studies of the resonance
  structure in \mbox{\decay{\Dz}{\KS\Kpm\pimp}} decays}},
  }{}\href{https://doi.org/10.1103/PhysRevD.93.052018}{Phys.\ Rev.\
  \textbf{D93} (2016) 052018},
  \href{http://arxiv.org/abs/1509.06628}{{\normalfont\ttfamily
  arXiv:1509.06628}}\relax
\mciteBstWouldAddEndPuncttrue
\mciteSetBstMidEndSepPunct{\mcitedefaultmidpunct}
{\mcitedefaultendpunct}{\mcitedefaultseppunct}\relax
\EndOfBibitem
\bibitem{LHCb-PAPER-2016-003}
LHCb collaboration, R.~Aaij {\em et~al.},
  \ifthenelse{\boolean{articletitles}}{\emph{{Measurement of \CP observables in
  \mbox{\decay{\Bpm}{\D \Kpm}} and \mbox{\decay{\Bpm}{\D\pipm}} with two- and
  four-body \D decays}},
  }{}\href{https://doi.org/10.1016/j.physletb.2016.06.022}{Phys.\ Lett.\
  \textbf{B760} (2016) 117},
  \href{http://arxiv.org/abs/1603.08993}{{\normalfont\ttfamily
  arXiv:1603.08993}}\relax
\mciteBstWouldAddEndPuncttrue
\mciteSetBstMidEndSepPunct{\mcitedefaultmidpunct}
{\mcitedefaultendpunct}{\mcitedefaultseppunct}\relax
\EndOfBibitem
\end{mcitethebibliography}

\newpage
\centerline
{\large\bf LHCb collaboration}
\begin
{flushleft}
\small
R.~Aaij$^{31}$,
C.~Abell{\'a}n~Beteta$^{49}$,
T.~Ackernley$^{59}$,
B.~Adeva$^{45}$,
M.~Adinolfi$^{53}$,
H.~Afsharnia$^{9}$,
C.A.~Aidala$^{80}$,
S.~Aiola$^{25}$,
Z.~Ajaltouni$^{9}$,
S.~Akar$^{66}$
% ...

{\footnotesize \it

% ...
$ ^{9}$Universit{\'e} Clermont Auvergne, CNRS/IN2P3, LPC, Clermont-Ferrand, France\\
% ...
$ ^{25}$INFN Sezione di Milano, Milano, Italy\\
% ...
$ ^{31}$Nikhef National Institute for Subatomic Physics, Amsterdam, Netherlands\\
% ...
$ ^{45}$Instituto Galego de F{\'\i}sica de Altas Enerx{\'\i}as (IGFAE), Universidade de Santiago de Compostela, Santiago de Compostela, Spain\\
% ...
$ ^{49}$Physik-Institut, Universit{\"a}t Z{\"u}rich, Z{\"u}rich, Switzerland\\
% ...
$ ^{53}$H.H. Wills Physics Laboratory, University of Bristol, Bristol, United Kingdom\\
% ...
$ ^{59}$Oliver Lodge Laboratory, University of Liverpool, Liverpool, United Kingdom\\
% ...
$ ^{66}$Los Alamos National Laboratory (LANL), Los Alamos, United States\\
$ ^{67}$Syracuse University, Syracuse, NY, United States\\
% ...
$ ^{80}$University of Michigan, Ann Arbor, United States, associated to $^{67}$\\
\bigskip
% ...
\medskip
}
\end{flushleft}

\end{document}